\documentclass[12pt]{article}       
\usepackage{amsmath}
\usepackage{amsfonts}               

\DeclareMathSizes{12}{11}{8}{6}

\DeclareMathOperator*{\Pf}{Pf}
\DeclareMathOperator*{\erf}{erf}
\DeclareMathOperator*{\erfc}{erfc}
\DeclareMathOperator*{\Tr}{Tr}
\DeclareMathOperator*{\Ai}{Ai}
\DeclareMathOperator*{\sgn}{sgn}
\DeclareMathOperator*{\intpart}{int}
\DeclareMathOperator*{\inteven}{int_{even}}
\DeclareMathOperator*{\intodd}{int_{odd}}
\DeclareMathOperator*{\Aid}{Aid}

\newcommand{\nn}{\nonumber}

\newcommand{\e}{\mbox{e}}

\newcommand{\mcF}{\mathcal{F}}
\newcommand{\mcK}{\mathcal{K}}
\newcommand{\mcO}{\mathcal{O}}

\newcommand{\RR}{\mathbb{R}}
\newcommand{\CC}{\mathbb{C}}

\newcommand{\intC}{\int_{\CC}}	
\newcommand{\intR}{\int_{\RR}}	

\newcommand{\dsq}{d^{\hspace{0.5pt}2}}

\newcommand{\iim}{\Im m\,}
\newcommand{\rre}{\Re e\,}

\newcommand{\be}{\begin{equation}}
\newcommand{\ee}{\end{equation}}

\begin{document}

\renewcommand\baselinestretch{1.2}
\baselineskip=18pt plus1pt		
\newcommand{\sect}[1]{\setcounter{equation}{0}\section{#1}}
\renewcommand{\theequation}{\thesection.\arabic{equation}}

\author{~\\{\sc G.~Akemann}$^1$ and {\sc M.~J.~Phillips}$^{2}$
\\~\\$^1$Department of Physics,
Bielefeld University,\\
Postfach 100131,
D-33501 Bielefeld, Germany
\\~\\
$^2$School of Mathematical Sciences, Queen Mary University of London,
\\ London E1 4NS, United Kingdom}
\date{}
\title{The Interpolating Airy Kernels for the $\beta=1$ and $\beta=4$ Elliptic Ginibre Ensembles}
\maketitle
\vfill
\begin{abstract}
We consider two families of non-Hermitian Gaussian random matrices,
namely the elliptical Ginibre ensembles of asymmetric $N$-by-$N$ matrices
with Dyson index $\beta=1$ (real elements) and with $\beta=4$ (quaternion-real elements).
Both ensembles have already been solved for finite $N$ using the method of skew-orthogonal polynomials,
given for these particular ensembles in terms of Hermite polynomials in the complex plane.
In this paper we investigate the microscopic weakly non-Hermitian large-$N$ limit of each ensemble
in the vicinity of the largest or smallest real eigenvalue.
Specifically, we derive the limiting matrix-kernels for each case,
from which all the eigenvalue correlation functions can be determined.
We call these new kernels the ``interpolating'' Airy kernels,
since we can recover -- as opposing limiting cases -- not only the well-known Airy kernels for the Hermitian ensembles,
but also the complementary error function and Poisson kernels
for the maximally non-Hermitian ensembles at the edge of the spectrum.
Together with the known interpolating Airy kernel for $\beta=2$, which we rederive here as well,
this completes the analysis of all three elliptical Ginibre ensembles in the microscopic scaling limit
at the spectral edge.

\end{abstract}

\thispagestyle{empty}
\newpage

\sect{Introduction}\label{intro}

Non-Hermitian Random Matrix Theory (RMT) is of considerable interest for various reasons,
including its many interesting applications in physics and elsewhere.
We refer to \cite{FyoSom03,AZ,Ake07} for review articles and references.

In this paper, we consider the elliptical Ginibre ensembles of $N$-by-$N$ matrices
with real, complex or quaternion-real Gaussian matrix elements,
labelled by the Dyson index $\beta=1,2$ and $4$ respectively.
These ensembles incorporate a non-Hermiticity parameter
that allows us to interpolate between the classical Gaussian ensembles (GOE, GUE and GSE)
and the corresponding maximally non-Hermitian Ginibre ensembles (GinOE, GinUE and GinSE).
The eigenvalue correlation functions can be expressed in terms of the
parameter-dependent kernels of each ensemble.
See \cite{BKHJS} for a detailed overview of these ensembles.

We can determine a number of distinct large-$N$ limits of these elliptical ensembles.
The limits can be macroscopic or microscopic,
and taken in different regions of the spectrum (e.g.\ in the bulk or at the edge),
depending on how the eigenvalues are scaled and shifted.
The limits can also be either strongly non-Hermitian,
where the degree of non-Hermiticity remains constant as the large-$N$ limit is taken,
or weakly non-Hermitian,
where the degree of non-Hermiticity becomes vanishingly small.
In the latter case, a further scaling of the eigenvalues must be performed,
to obtain a limit that is distinct from the Hermitian case.

Not all of these large-$N$ limits have been explored to date,
and the aim of this paper is to complete the missing cases
by determining the microscopic, weakly non-Hermitian limits
for the $\beta=1$ and $\beta=4$ ensembles,
in the vicinity of the largest (or smallest) real eigenvalue.
These new results were previously announced in \cite{APh}.
The $\beta=2$ case was already known \cite{Bender},
and, in fact, it has been shown \cite{ABe} that the chiral $\beta=2$ ensemble also has the same limit.

These weakly non-Hermitian kernels may also be described as ``interpolating'',
since, for each $\beta$,
it is possible to recover both the Hermitian (Airy)
and the strongly non-Hermitian complementary error function and Poisson kernels
by taking opposing limits.
In this paper we explicitly check that these limits agree with the known results.
Of course, in the Hermitian limit the distributions of the largest eigenvalue
for each of $\beta=1,2$ and $4$ can be written in terms
of the Fredholm determinant (or Pfaffian) of the corresponding Airy kernel \cite{Peter}.
These so-called Tracy-Widom distributions
can also be expressed (see \cite{TW}) in terms of the solution of a certain Painlev\'e equation.
To date, little is known about how these constructions might generalise
to the non-Hermitian case.
A first step was made in \cite{Bender} with an explicit Fredholm determinant
construction based on the interpolating Airy kernel for $\beta=2$.
However, no relation to non-linear equations such as the Painlev\'e equations
is apparent in this form.

On the other hand, at maximal non-Hermiticity it has been shown (see \cite{Rider})
for $\beta=2, 4$ and most recently $\beta=1$
that the eigenvalue with the largest modulus
follows the Gumbel distribution.
Furthermore, the real part of the eigenvalue with
largest real part also obeys the Gumbel distribution in the strongly non-Hermitian limit,
at least for the $\beta=2$ case, see \cite{Bender}.
The latter result was derived by taking the strongly non-Hermitian limit
of the Fredholm determinant of the interpolating Airy kernel.
We extend this result to the $\beta=4$ and $\beta=1$ ensembles in the present paper.

As an aside we mention that in \cite{Kurt} a transition between the
Tracy-Widom and Gumbel distributions has been found;
however, this was for an ensemble where the eigenvalues always remain real.

One of the most striking features of RMT,
and one which directly underlies its physical applicability,
is that certain large-$N$ limits
turn out to be independent of the specific details of the distribution of the matrix elements.
This is a property known as universality.
Whilst it is now very well understood under which general conditions universality holds for Hermitian RMT,
universality for the non-Hermitian ensembles has been explored less
(see \cite{BKHJS} for an overview as well as \cite{ameur,berman,taovu}),
in particular the weakly non-Hermitian limit (see \cite{FKS2,FKS,Ake02}).
This is perhaps quite surprising,
since all the weakly non-Hermitian kernels can be expressed \cite{APh}
as one-parameter deformations of the corresponding Airy, sine or Bessel\footnote{The Bessel kernels arise from the Gaussian chiral ensembles.}
kernels for real eigenvalues,
leading to the conjecture \cite{APh} (and subsequent research programme) that universality holds
for all the weakly non-Hermitian kernels listed therein.
In fact, further evidence is provided by the fact that the same interpolating kernels
also appear when interpolating between two different \textit{Hermitian} ensembles,
see \cite{ADOS,FNH,Macedo}.

This paper is organised as follows. In the next section, we define the elliptical ensembles
for $\beta=1,2$ and $4$, restating the known solutions at finite $N$.
In Section \ref{KAib2}, we rederive the known weakly non-Hermitian results for the $\beta=2$ case
in the vicinity of the largest real eigenvalue,
in order to illustrate our techniques.
The new results for $\beta=4$ and $\beta=1$ then follow in Sections \ref{KAib4} and \ref{KAib1} respectively,
showing an increasing degree of difficulty.
Each section ends with a cross-check of the Hermitian and strongly non-Hermitian limits,
by sending the deformation parameter to zero or infinity.
Several technical details of these limits are postponed until the appendices.
We also show for the $\beta=2$ case that the weakly non-Hermitian kernel
in the bulk can be recovered from the interpolating Airy kernel\footnote{This
can be demonstrated for the other two ensembles as well,
although rigorous proofs are beyond the scope of this paper.}.
Finally, our conclusions are presented in Section \ref{conclusions}.

\sect{The matrix ensembles at finite $N$}\label{finiteN}

The partition function of the three families of elliptic Ginibre ensembles labelled by Dyson index $\beta=1,2,4$ is defined as
\be
{\cal Z}_{N}^{(\beta)}(\tau)\equiv \int dJ\exp\left[-\frac{1}{1-\tau^2}
\Tr\Big(
JJ^\dag-\frac{\tau}{2}(J^2+J^{\dag\,2})\Big)\right]=
\int dH\,dA \exp\left[-\frac{\Tr H^2}{1+\tau}+
\frac{\Tr A^2}{1-\tau}\right].
\label{GinbE}
\ee
Here the elements of the $N\times N$ matrix $J$ are chosen to be real ($\beta=1$), complex ($\beta=2$), or quaternion-real ($\beta=4$), without any further symmetry constraints. When splitting the matrix $J$ into its Hermitian and anti-Hermitian parts, $J=H+A$, it can be seen that this
non-Hermitian matrix model is equivalent to a Gaussian two-matrix model. The parameter $\tau\in[0,1)$ allows us to interpolate between the Ginibre ensembles at maximal non-Hermiticity when $\tau=0$, and the classical Wigner-Dyson ensembles in the Hermitian limit $\tau\to1$.

The eigenvalue representations of these partition functions are derived
using a Schur decomposition for $\beta=2$ and $\beta=4$ \cite{Ginibre} and a $QR$ decomposition for $\beta=1$ \cite{Lehmann-Sommers,Edelman}. The resulting reduction to integrals over the complex eigenvalues (and real eigenvalues for $\beta=1$) of the matrix $J$ takes the following forms.
For $\beta=2$ we have
\be
{\cal Z}_{N}^{(\beta=2)}(\tau)=c_{N}^{(\beta=2)}
\prod_{j=1}^N\int_{\mathbb C}\dsq z_j\,w^{(\beta=2)}(z_j)\
|\Delta_N(\{z\})|^2,
\label{ZGinUE}
\ee
where $\Delta_N(\{z\}) \equiv \det[z_i^{j-1}]_{1\le i,j \le N} =\prod_{1\leq j<k\leq N}(z_k-z_j)$
is the Vandermonde determinant,
$c_{N}^{(\beta=2)}$ is a known normalisation factor,
and we have a real-valued weight function given in the complex plane by
\be
w^{(\beta=2)}(z)
= \exp\left[-\frac{1}{1-\tau^2}
\Big(
|z|^2-\frac{\tau}{2}(z^2+z^{*\,2})\Big)\right] =
\exp\left[- \, \frac{x^2}{1+\tau} - \frac{y^2}{1-\tau}\right] =  w^{(\beta=4)}(z)\ ,
\label{weightb24}
\ee
in which $x=\rre(z)$ and $y=\iim(z)$.
We will encounter the same weight function for the $\beta=4$ ensemble below.
For convenience, we suppress the $\tau$-dependency of $c_{N}^{(\beta)}$ and $w^{(\beta)}(z)$ in our notation.

For $\beta=4$ we have
\begin{align}
{\cal Z}_{N}^{(\beta=4)}(\tau)
&= c_{N}^{(\beta=4)}
\prod_{j=1}^N\int_{\mathbb C}\dsq z_j\,|z_j-z_j^*|^2w^{(\beta=4)}(z_j)
\prod_{1\leq j<k\leq N}|z_k-z_j|^2|z_k-z_j^*|^2\nn\\
&= c_{N}^{(\beta=4)}
\prod_{j=1}^{2N}\int_{\mathbb C}\dsq z_j
\prod_{k=1}^N\mcF^{(\beta=4)}(z_{2k-1},z_{2k})\
\Delta_{2N}(\{z\})\ .
\label{ZGinSE}
\end{align}
In the first line we give the Jacobian as computed by Ginibre
\cite{Ginibre},
from which it can be seen that the eigenvalues are repelled from the real axis.
In the second line, following \cite{Kanzieper_2002_02,BKHJS}, we write the integral
over all $2N$ eigenvalues (i.e.\ including every eigenvalue and its perfectly correlated complex conjugate)
with a single Vandermonde determinant of size $2N$
and a product over an anti-symmetric bivariate weight function defined as
\be
\label{beta_eq_4_F_def}
\mcF^{(\beta=4)}(z_1,z_2)
= \sqrt{w^{(\beta=4)}(z_1)w^{(\beta=4)}(z_2)} \, (z_1 - z_2) \, \delta^{(2)}(z_1 - z_2^*) \ ,
\ee
where $\delta^{(2)}(z)\equiv \delta(x)\delta(y)$ is the 2-dimensional Dirac delta function.
The second form in eq.\ (\ref{ZGinSE}) emphasises the similarity to the $\beta=1$ ensemble below, and it is easy to see that the two formulations are equivalent by using the results in \cite{Kanzieper_2002_02}.

Finally we give the result for $\beta=1$
where we restrict ourselves to even $N$,
as we will do throughout this paper:
\be
{\cal Z}_{N}^{(\beta=1)}(\tau)=c_{N}^{(\beta=1)}
\prod_{j=1}^N\int_{\mathbb  C} \dsq z_j
\prod_{k=1}^{N/2} {\cal F}^{(\beta=1)}(z_{2k-1},z_{2k})\
\Delta_N(\{z\})\ .
\label{ZGinOE}
\ee
The $\beta=1$ partition function differs from the $\beta=4$ case
both in the form of the anti-symmetric bivariate weight
function\footnote{In \cite{BKHJS} the $\beta=4$ case was cast into the same form as for $\beta=1$.}
\be
\label{beta_eq_1_F_def}
\mcF^{(\beta=1)}(z_1,z_2) = w^{(\beta=1)}(z_1)w^{(\beta=1)}(z_2)\Big\{ 2i\delta^{(2)}(z_1-z_2^*)\sgn(y_1) + \delta(y_1)\delta(y_2)\sgn(x_2-x_1) \Big\},
\ee
where the function $\sgn(x)$ denotes the sign function,
and in the weight function itself
\begin{align}
w^{(\beta=1)}(z) &= \exp\left[ - \, \frac{x^2}{2(1+\tau)} + \frac{y^2}{2(1+\tau)} \right] \left[ \erfc\left( \sqrt{\frac{2}{1-\tau^2}}\, |y| \right) \right]^{1/2} \label{weightb1} \\
 &= \exp\left[ \frac{y^2}{1-\tau^2} \right] \, \left[ \erfc\left( \sqrt{\frac{2}{1-\tau^2}} \,|y| \right) \right]^{1/2} \, \sqrt{w^{(\beta=2)}(z)} . \label{weightb1_vs_b2}
\end{align}
The bivariate weight function eq.\ (\ref{beta_eq_1_F_def}) contains two parts,
leading to, respectively, complex conjugate eigenvalue pairs and real eigenvalues.
It should be noted that the integrand in the partition function is not always positive,
and so is not a true joint probability density function (jpdf).
One must apply a symmetrisation procedure
when determining observables such as correlation functions.

We now present the known results for the correlation functions of complex eigenvalues.
For $\beta=2$  we use the partition function
${\cal Z}_{N}^{(\beta=2)}(\tau)=c_{N}^{(\beta=2)}
\prod_{j=1}^N\int_{\mathbb  C} \dsq z_j\, {\cal P}^{(\beta=2)}_{\text{jpdf}}(\{z\})$
and corresponding jpdf ${\cal P}^{(\beta=2)}_{\text{jpdf}}(\{z\})$
to define the $k$-point correlation functions as
\be
R_{k,N}^{(\beta=2)}(z_1,\ldots,z_k)\equiv\frac{N!}{(N-k)!}
\frac{1}{{\cal Z}_{N}^{(\beta=2)}(\tau)}
\prod_{j=k+1}^N\int_{\mathbb  C} \dsq z_j\, {\cal P}^{(\beta=2)}_{\text{jpdf}}(\{z\})\ .
\label{Rkb2def}
\ee
These can be computed explicitly \cite{FKS} as
\be
R_{k,N}^{(\beta=2)}(z_1,\ldots,z_k)
 = \det_{i,j=1,\ldots,k}\left[\mcK_N^{(\beta=2)}(z_i,z_j^*)\right]
\ ,
\label{Rkb2}
\ee
where $\mcK_N^{(\beta=2)}(z_1,z_2)$ is the kernel associated with the weight function $w^{(\beta=2)}(z)$
in the complex plane,
written most simply in terms of the corresponding orthogonal polynomials.
We will give the kernel explicitly in the next section.
Note that for $k=N$ we see that the jpdf itself can be written as a determinant,
proven directly by algebraic manipulation of eq.\ (\ref{ZGinUE}).

For $\beta=1,4$, the correlation functions are given in terms of the Pfaffian of a matrix-kernel.
For $\beta=4$, they follow from \cite{Kanzieper_2002_02}
and for $\beta=1$ from \cite{HJS-Gin}:
\begin{align}
R_{k,N}^{(\beta=1,4)}(z_1,\ldots,z_k) &= \Pf_{i,j=1,\ldots,k}
\left[\left(
\begin{array}{cc}
\hat{\mcK}_{\alpha N}^{(\beta=1,4)}(z_i,z_j)
&-G_{\alpha N}^{(\beta=1,4)}(z_i,z_j)\\
G_{\alpha N}^{(\beta=1,4)}(z_j,z_i)
&-W_{\alpha N}^{(\beta=1,4)}(z_i,z_j)\\
\end{array}
\right)\right]\ ,
\label{Rkb4}
\end{align}
where $\alpha=1$ for $\beta=1$, and $\alpha=2$ for $\beta=4$.
(We introduce $\alpha$ in this way because, in our convention, the subscript label on the kernel elements
always counts the total number of eigenvalues, and in the $\beta=4$ case, an $N$-by-$N$ matrix
of quaternion-valued elements has $2N$ complex-valued eigenvalues.)
Here we have also defined the elements of the matrix-kernel as
\begin{align}
G_N^{(\beta=1,4)}(z_1,z_2) & = - \, \intC \dsq z\,\hat{\mcK}^{(\beta=1,4)}_N(z_1,z)\,\mcF^{(\beta=1,4)}(z,z_2) \ ,\label{GN_def} \\
W_N^{(\beta=1,4)}(z_1,z_2) & = - \, \intC \dsq z \, \mcF^{(\beta=1,4)}(z_1,z)\,G_N^{(\beta=1,4)}(z,z_2) - \mcF^{(\beta=1,4)}(z_1,z_2)\ . \label{WN_def}
\end{align}
The function $\hat{\mcK}^{(\beta=1,4)}_N(z_1,z_2)$ above denotes the pre-kernel
associated with the bivariate weight function $\mcF^{(\beta=1,4)}(z_1,z_2)$,
written most conveniently in terms of the corresponding skew-orthogonal polynomials
which we will give explicitly in the following sections.
For $\beta=1$ and odd $N$ we refer to \cite{HJSW,FM}.

Note that for $\beta=1$ and $\beta=4$, the $k$-point correlation functions (for $k \ge 2$) contain so-called
``contact terms'', corresponding to the perfect correlation between a complex eigenvalue $z$
and its complex conjugate $z^*$.
It is easy to remove such contact terms, should this be desired:
for $\beta=4$, for example, one can redefine
$W_N^{(\beta=4)}(z_1,z_2)\to W_N^{(\beta=4)}(z_1,z_2)+\mcF^{(\beta=4)}(z_1,z_2)$,
leading to the form first derived in \cite{Kanzieper_2002_02}.
A similar procedure can be performed for the $\beta=1$ case.

In the following sections we will analyse the microscopic edge scaling limits of the kernels $\mcK_{N}^{(\beta)}(z_1,z_2)$,
which are given in terms of Hermite polynomials in the complex plane for each $\beta=1,2,4$,
as well as of its companions $G_N^{(\beta)}(z_1,z_2)$ and $W_N^{(\beta)}(z_1,z_2)$ for $\beta=1,4$.
The result in the next section for $\beta=2$ is not new,
but our alternative (and straightforward) rederivation of the results in \cite{Bender,ABe}
will be a very instructive preparation for the more complicated cases $\beta=4$ and $\beta=1$
in the subsequent sections.

\sect{The $\beta=2$ interpolating Airy kernel revisited}\label{KAib2}

In this section we rederive the large-$N$ microscopic weakly non-Hermitian asymptotic limit
of the $\beta=2$ kernel $\mcK_{N}^{(2)}(z_1,z_2)$ in the vicinity of the largest or smallest real eigenvalue
(i.e.\ the edge scaling limit).
The same result was derived rigorously in \cite{Bender} at the largest real eigenvalue,
and we reproduce here the real integral representation of the interpolating Airy kernel given in \cite{ABe}.

There are several reasons why we rederive this result here.
First, our proof is very short and easy to follow, although we omit some technical details.
It is actually simplest for $\beta=2$, and serves as a preparation for the derivations
for $\beta=4$ and $\beta=1$ which will show an increasing degree of complexity.
Last but not least we are able to demonstrate why our result also applies in the vicinity
of the smallest real eigenvalue.

By using eq.\ (\ref{Rkb2}), all the correlations functions can be expressed at finite $N$
in terms of the kernel of orthogonal polynomials (OP), which is given \cite{FKS} by
\be
\mcK_N^{(2)}(z_1,z_2) = \sqrt{w^{(2)}(z_1)w^{(2)}(z_2)} \,\, \sum_{j=0}^{N-1} \frac{1}{c_j} p_j(z_1)p_j(z_2)\ .
\label{KNb2}
\ee
The OP are in monic normalisation, and are orthogonal in the complex plane
with respect to the weight function $w^{(2)}(z)$:
\be
\int_{\mathbb  C} \dsq z\ w^{(2)}(z)\ p_j(z)\ p_k(z^*)=c_j\delta_{jk}\ .
\ee
Specifically, the polynomials and (squared) norms are given by
\begin{align}
p_j(z) &= \left( \frac{\tau}{2} \right)^{j/2} \, H_j\left( \frac{z}{\sqrt{2\tau}} \right)\ ,
\label{OPb2}\\
c_j &= \pi \, j! \, (1-\tau^2)^{1/2}\ ,
\label{normb2}
\end{align}
where $H_j(z)$ is the physicists' Hermite polynomial\footnote{Note that, in contrast to the present paper, the probabilists' Hermite polynomials $\text{He}_j(z)$ were used in \cite{FKS}.} of degree $j$.
This was shown independently in \cite{Hermite} and \cite{PdF}. See also \cite{BKHJS} for a short proof.

In the macroscopic large-$N$ limit where $\tau$ is kept fixed,
the mean spectral density defined as
$\displaystyle \rho(z) \equiv \lim_{N\rightarrow\infty} R_{1,N}^{(2)}(\sqrt{N}\,z)$, with $R_{1,N}^{(2)}(z)=\mcK_N^{(2)}(z,z^*)$,
is constant on an ellipse \cite{FKS}
with axes $1+\tau\geq|\rre (z)|$ and $1-\tau\geq|\iim (z)|$.
Conversely, we can define the weakly non-Hermitian microscopic large-$N$ limit as
\be
\lim_{N\to\infty}N^{1/3}(1-\tau)=\sigma^2\ ,
\label{weaklim}
\ee
where now $\sigma$ is fixed (and so $\tau\rightarrow 1$),
and simultaneously magnify the vicinity of either the largest or
the smallest real eigenvalue at $\pm(1+\tau)\sqrt{N}$ as
\be
z=\pm\Big((1+\tau) \sqrt{N} + \frac{Z}{N^{1/6}}\Big)\ ,
\label{edgelim}
\ee
for fixed microscopic coordinate $Z=X+iY\in\mathbb{C}$.
Under this scaling we can rederive following interpolating Airy kernel \cite{Bender,ABe}:
\begin{align}
\label{GinUE_edge_final}
\mcK^{(2)}_{\text{Ai}}(Z_1,Z_2) &\equiv \lim_{\substack{N\rightarrow\infty \\ \tau\to1}} \frac{1}{N^{1/3}} e^{-iN^{1/3}(Y_1+Y_2)}\mcK_N^{(2)}\left(
\pm\Big((1+\tau) \sqrt{N} + \frac{Z_1}{N^{1/6}}\Big),\pm
\Big((1+\tau) \sqrt{N} + \frac{Z_2}{N^{1/6}}\Big)
\right) \nn \\
 &= \frac{1}{\sigma\sqrt{\pi}} \, \exp\left[ - \, \frac{Y_1^2+Y_2^2}{2\sigma^2} + \frac{\sigma^6}{6} + \frac{\sigma^2(Z_1+Z_2)}{2} \right]
\nn \\
& \qquad \times
\int_0^{\infty} dt \, \e^{\sigma^2 t} \Ai\left( Z_1 + \frac{\sigma^4}{4} + t \right)\, \Ai\left( Z_2 + \frac{\sigma^4}{4} + t \right).
\end{align}
The effective redefinition of the finite-$N$ kernel to include an $N$-dependent phase factor has no effect
on the $k$-point correlation functions, since it will cancel when the determinant in eq.\ (\ref{Rkb2}) is evaluated.
However, this factor ensures that the kernel itself has a well-defined limit as $N\rightarrow\infty$.

As a particular example, the microscopic density is given by
\be
\label{microb2}
{R}^{(2)}_{1,\text{Ai}}(Z)=\mcK^{(2)}_{\text{Ai}}(Z,Z^*)=
\frac{1}{\sigma\sqrt{\pi}} \, \exp\left[ - \, \frac{Y^2}{\sigma^2} + \frac{\sigma^6}{6} + \sigma^2X \right]
\int_0^{\infty} dt \, \e^{\sigma^2 t} \, \left|\Ai\left( Z + \frac{\sigma^4}{4} + t \right)\right|^2\ .
\ee

The case with negative signs in eq.\ (\ref{edgelim}),
corresponding to magnifying the region around the smallest real eigenvalue,
can easily be mapped to that with positive signs due to the following observation:
Since the weight function $w^{(2)}(z)$ in eq.\ (\ref{weightb24}) is even in both its real and imaginary parts,
and the Hermite polynomials appearing twice in the kernel have parity $H_j(-z)=(-)^jH_j(z)$,
we obtain
\be
\mcK_N^{(2)}(-z_1,-z_2)=\mcK_N^{(2)}(+z_1,+z_2)\ .
\ee
Hence we can focus just on the positive signs in eq.\ (\ref{GinUE_edge_final}), as in \cite{Bender,ABe}.

Our derivation is based on the well known asymptotic limit of the Hermite polynomials
at fixed $Z$ which follows from \cite{PlanRot,Szego}:
\be
\label{Hermite_2_Airy_asymptotic}
\lim_{n\rightarrow\infty} \left\{ \frac{1}{(2\pi)^{1/4}2^{n/2}} \, \frac{n^{1/12}}{\sqrt{n!}} \, \e^{-z^2/2}
H_n(z)\right\}
= \Ai(Z)\ ,\ \
\mbox{where} \ \ \ z=\sqrt{2n} + \frac{Z}{\sqrt{2}\,n^{1/6}}\ ,
\ee
and where $\Ai(z)$ is the Airy function.
This limit is known to hold for complex $z$.
We already see from this that it will be essential to consider the large-$N$ limit of
the product of the weight function $\sqrt{w^{(2)}(z)}$ and the OP with corresponding argument $p_j(z)$ in eq.\ (\ref{KNb2}) together, rather than each factor separately.

Because the standard  Christoffel-Darboux formula does not apply to OP in the complex
plane\footnote{But see \cite{Bohigas} for an alternative approach to this issue.},
we must work directly with the sum in eq.\ (\ref{KNb2}).
First, we change the summation index from $j$ to $t$ as follows:
\be
\mcK_N^{(2)}(z_1,z_2) =
\sqrt{w^{(2)}(z_1)w^{(2)}(z_2)} \,\, \sum_{\substack{t=\Delta t \\ \Delta t = N^{-1/3}}}^{t = N\Delta t = N^{2/3}} \frac{1}{c_j} p_j(z_1) p_j(z_2),
\ee
where
\be
j \equiv j(t,N) = N - tN^{1/3},
\ee
and $t$ takes discrete values $\Delta t, 2\Delta t, \ldots, N\Delta t$ where $\Delta t = N^{-1/3}$.
If we now allow $t$ to be continuous in the range $0 \le t \le N^{2/3}$,
rather than taking only integer multiples of $\Delta t$, and redefine $j$ as
\be
j \equiv j(t,N) = \intpart(N-tN^{1/3}),
\label{jN}
\ee
where the function $\intpart(x)$ denotes the integer part of $x$,
then we can write the sum as an integral over a step function as follows:
\begin{align}
\mcK_N^{(2)}(z_1,z_2) &= N^{1/3} \,
\sqrt{w^{(2)}(z_1)w^{(2)}(z_2)} \,\, \int_0^{N^{2/3}} dt \,  \frac{1}{c_j} p_j(z_1) p_j(z_2) \nn \\
 &= N^{1/3} \int_0^{\infty} dt \,  \frac{1}{c_j} \left(\sqrt{w^{(2)}(z_1)} p_j(z_1)\right) \left( \sqrt{w^{(2)}(z_2)} p_j(z_2) \right) \, \Theta(j).
\end{align}
In the second step here, we introduced the Heaviside theta function
\be
\Theta(x) = \begin{cases} 1 & \text{if $x > 0$} \\ \tfrac12 & \text{if $x=0$} \\ 0 & \text{if $x < 0$}  \end{cases}
\ee
so that we could change the upper limit of the integral to be independent of $N$, as required in the following.
We reiterate that this representation is still exact at finite $N$.

We wish to take the large-$N$ limit
\be
\mcK^{(2)}_{\text{Ai}}(Z_1,Z_2) \equiv \lim_{\substack{N\rightarrow\infty \\ \tau\to1}}
\frac{1}{N^{1/3}}
e^{-iN^{1/3}(Y_1+Y_2)}
\mcK_N^{(2)}(z_1,z_2)
\ee
under the scalings eqs.\ (\ref{weaklim}) and (\ref{edgelim}).
In fact, we can take the limit inside the integral,
a step that can be fully justified by invoking the Dominated Convergence Theorem.
Our task is therefore to determine the large-$N$ behaviour of the individual factors
in the integrand at fixed $t$.
For the reciprocal of the squared norm, we write at fixed $t$ and large $N$ (and hence large $j$)
\begin{align}
\frac{1}{c_j}
 & = \frac{1}{\pi\,j!\,(1+\tau)^{1/2}(1-\tau)^{1/2}}
 \sim \frac{N^{1/6}}{\sqrt{2} \, \pi\,j!\, \sigma}
 \sim \frac{1}{\sqrt{\pi}\,\sigma} \left( \frac{j^{1/12}}{(2\pi)^{1/4}\sqrt{j!}} \right)^2,
\end{align}
and so
\begin{align}
\mcK^{(2)}_{\text{Ai}}(Z_1,Z_2)
&=
\frac{1}{\sqrt{\pi}\,\sigma} \int_0^{\infty} dt \,
\lim_{\substack{N\rightarrow\infty \\ \tau\to1}}
\Big( e^{-iN^{1/3}Y_1} h_j(z_1) \Theta(j) \Big)
\lim_{\substack{N\rightarrow\infty \\ \tau\to1}}
\Big( e^{-iN^{1/3}Y_2} h_j(z_2) \Theta(j) \Big),
\label{Ksymp}
\end{align}
where
\be\label{beta_eq_2_hj_def}
h_j(z) \equiv \tau^{j/2} \, \frac{j^{1/12}}{(2\pi)^{1/4} 2^{j/2}\sqrt{j!}} \, \sqrt{w^{(2)}(z)} \, H_j\left( \frac{z}{\sqrt{2\tau}} \right)
\ee
and we emphasise that the limit is taken at fixed $t$, $Z$ and $\sigma$,
with $j$, $z$ and $\tau$ being dependent on these, and on $N$.

In order to apply eq.\ (\ref{Hermite_2_Airy_asymptotic}) we need to implement carefully the scalings eqs.\ (\ref{weaklim}) and (\ref{edgelim}) as well as the change of variables eq.\ (\ref{jN}).
We begin with the argument of the Hermite polynomials.
First, we expand ${1/\sqrt{\tau}}$ in (decreasing) powers of $N$ up to the order that we need later:
\be
\frac{1}{\sqrt{\tau}}
= \left( 1 - \frac{\sigma^2}{N^{1/3}} \right)^{-1/2} 
= 1 + \frac{\sigma^2}{2N^{1/3}} + \frac{3\sigma^4}{8N^{2/3}}
+ \frac{5\sigma^6}{16N} + \mcO(N^{-4/3}).
\ee
Using this we obtain for the argument
\be
\frac{z}{\sqrt{2\tau}}
= \frac{1}{\sqrt{2\tau}} \, \left( (1+\tau)\sqrt{N} + \frac{Z}{N^{1/6}} \right)
= \sqrt{2N} + \frac{Z+\sigma^4/4}{\sqrt{2}\,N^{1/6}}+\frac{2\sigma^2Z + \sigma^6}{4\sqrt{2N}} +\mcO(N^{-5/6}).
\label{zexp}
\ee
Now we have to write this in terms of the degree $j$ $(= \intpart(N - t N^{1/3})$) of the Hermite polynomial.
By adding and subtracting $\sqrt{2j}$ we have
\begin{align}
\frac{z}{\sqrt{2\tau}} &=
\sqrt{2N} + \frac{Z+\sigma^4/4}{\sqrt{2}\,N^{1/6}} + \mcO(N^{-1/2}) \nn \\
 &=\sqrt{2j} + \sqrt{2N}\left( 1 - \sqrt{1 - \frac{t}{N^{2/3}}} \right)
+ \frac{Z+\sigma^4/4}{\sqrt{2}\,N^{1/6}} + \mcO(N^{-1/2}) \nn \\
 &= \sqrt{2j} + \frac{Z+t+\sigma^4/4}{\sqrt{2}\,j^{1/6}} + \mcO(N^{-1/2}) \ ,
 \end{align}
after expanding the square root.
By considering the coefficient of $j^{-1/6}$ in the argument of $H_j$,
we see that we need to make the substitution $Z \rightarrow Z+t+\sigma^4/4$
in the argument of the Airy function in the limit eq.\ (\ref{Hermite_2_Airy_asymptotic}).

Next we consider the weight function and expand
\begin{align}
\label{lim_weight}
\sqrt{w^{(2)}(z)}
 &= \exp\left[ - \, \frac{1}{2(1+\tau)} \left( (1+\tau)\sqrt{N} + \frac{X}{N^{1/6}} \right)^2 - \frac{1}{2(1-\tau)} \left( \frac{Y}{N^{1/6}} \right)^2 \right] \nn \\
 &= \exp \left[ -N + \frac{\sigma^2 N^{2/3}}{2} - N^{1/3}X - \frac{Y^2}{2\sigma^2} \right]
 \left( 1 +\mcO(N^{-1/3})\right).
\end{align}
Here we can actually drop all of the terms with negative powers of $N$
since these will disappear in the large-$N$ limit.

Recall from eq.\ (\ref{Hermite_2_Airy_asymptotic}) that,
in order to obtain an Airy function from a Hermite polynomial $H_j(u)$ (for arbitrary argument $u$)
in the large-$j$ limit,
we also require a factor $\exp[-u^2/2]$ (as well as other $j$-dependent factors).
To counterbalance this factor, therefore, we will be left with a factor $\exp[+u^2/2]$,
with $u=z/\sqrt{2\tau}$ expanded appropriately:
\begin{align}\label{lim_weight_residual}
\exp\left[ \frac{1}{2} \left( \frac{z}{\sqrt{2\tau}} \right)^2 \right]
&= \exp\left[ \frac{1}{2} \left( \sqrt{2N} + \frac{Z+\sigma^4/4}{\sqrt{2}\,N^{1/6}} + \frac{2\sigma^2 Z+\sigma^6}{4\sqrt{2N}} \right)^2 \right] \left( 1 + \mcO(N^{-1/3}) \right) \nn \\
 &= \exp \left[ N + N^{1/3}\left(X + iY + \frac{\sigma^4}{4}\right) + \frac{2\sigma^2 Z+\sigma^6}{4} \right]
 \left( 1 + \mcO(N^{-1/3}) \right).
\end{align}
Here we needed all the terms computed in eq.\ (\ref{zexp}),
but can now drop those terms with negative powers of $N$,
since they will vanish as $N\rightarrow\infty$.

Let us next consider the factor $\tau^{j/2}$. Inserting $j= \intpart(N - t N^{1/3}) = N - t N^{1/3} + \mcO(1)$ we have the behaviour at large $N$
\begin{align}\label{lim_tau_factor}
\tau^{j/2}
 &\sim \left( 1 - \frac{\sigma^2}{N^{1/3}} \right)^{N/2} \left( 1 - \frac{\sigma^2}{N^{1/3}} \right)^{-tN^{1/3}/2}
\sim \exp\left[ - \, \frac{\sigma^2N^{2/3}}{2} - \frac{\sigma^4N^{1/3}}{4} - \frac{\sigma^6}{6} + \frac{\sigma^2 t}{2}\right]\ ,
\end{align}
where in the final step we used the result that, as $N\rightarrow \infty$,
\be
\left( 1 - \frac{x}{N} \right)^{N^3}
= \exp\left[ N^3\log\left(1 -\frac{x}{N}\right)\right]
=\exp\left[ N^2x + \frac{Nx^2}{2} + \frac{x^3}{3} +\mcO(N^{-1})\right].
\ee
Finally, we have $\Theta(j)=1$ for any fixed $t$ at sufficiently high $N$. Therefore, on combining results eqs.\ (\ref{lim_weight}), (\ref{lim_weight_residual}) and (\ref{lim_tau_factor}), we find that we get many cancellations, giving as a total result
\be
\lim_{\substack{N\rightarrow\infty \\ \tau\to1}} \left( e^{-iN^{1/3}Y} h_j(z) \Theta(j) \right)
= \exp\left[ - \, \frac{Y^2}{2\sigma^2} + \frac{\sigma^6}{12} +\frac{\sigma^2 (Z+t)}{2} \right] \Ai\left( Z + \frac{\sigma^4}{4} + t \right).
\label{hjlim}
\ee
We have such contributions for both $Z=Z_1$ and $Z=Z_2$, and therefore we obtain the desired result
eq.\ (\ref{GinUE_edge_final}) from eq.\ (\ref{Ksymp}).

\subsection{Hermitian, strongly non-Hermitian and bulk limits}

As a first check we will take the Hermitian limit $\sigma\to0$ to show that the Airy kernel for real eigenvalues is recovered.
Since this has already been done in \cite{Bender,ABe} we can be brief.
Employing the following representation of the Dirac delta function,
\be
\lim_{\sigma\to0}\frac{\e^{-y^2/\sigma^2}}{\sqrt{\pi}\,\sigma} =\delta(y)\ ,
\label{deltadef1}
\ee
it is easy to see that
\be
\label{KAib2lim}
\lim_{\sigma\to0}\mcK^{(2)}_{\text{Ai}}(Z_1,Z_2)=
\sqrt{\delta(Y_1)\delta(Y_2)} \, \int_0^{\infty} dt \, \Ai(X_1 + t) \, \Ai(X_2 + t)\ ,
\ee
where the square roots of the delta functions are to be understood in the following sense:
When evaluating the determinant in eq.\ (\ref{Rkb2}) to determine the correlation functions,
we will always get pairs of such square root factors $\sqrt{\delta(Y_j)}\sqrt{\delta(-Y_j)}=\delta(Y_j)$.
Hence the $k$-point correlation function will contain the factor $\displaystyle \prod_{j=1}^k \delta(Y_j)$.
Recalling that these correlation functions were defined with respect to a two-dimensional measure in eq.\ (\ref{Rkb2def}), this then correctly reproduces the $k$-point correlation functions of real eigenvalues as the determinant of the well known Airy kernel, since
\be
\mcK_{\text{Ai,Herm}}^{(2)}(X_1,X_2) \equiv \int_0^{\infty} dt \, \Ai(X_1 + t) \, \Ai(X_2 + t)=\frac{\Ai(X_1)\Ai\hspace{0mm}'(X_2) - \Ai\hspace{0mm}'(X_1) \Ai(X_2)}{X_1 - X_2}\ .
\label{K2Herm_def}
\ee
In particular we obtain for the eigenvalue density
\be
\label{microb2real}
\lim_{\sigma\to0}
{R}^{(2)}_{1,\text{Ai}}(Z) = \delta(Y)\int_0^{\infty} dt \, \left[ \Ai(X + t) \right]^2=
\delta(Y)\left(\Ai\hspace{0mm}'(X)^2-X\Ai(X)^2\right)\ .
\ee

As a second check we take the strongly non-Hermitian limit by sending $\sigma\to\infty$
and rescaling the eigenvalues such that $\hat{Z}=Z/\sigma$ is kept fixed.
The rescaling of the kernel and of the eigenvalues maps the weakly non-Hermitian scaling limit with scale $N^{1/6}$
to the strongly non-Hermitian scaling limit with scale $N^{1/2}$.

It is simplest if we write the interpolating Airy kernel eq.\ (\ref{GinUE_edge_final})
in terms of the deformed Airy function $\Aid(z,\sigma)$, which we define in eq.\ (\ref{Aid_def}).
On using the appropriate limit of $\Aid(z,\sigma)$, see eq.\ (\ref{Aid_limit}),
and after interchanging the limit and the integral which can be justified,
it is then easy to show that the kernel has the limit
\begin{align}\label{strongdef}
\mcK^{(2)}_{\text{edge}}(\hat{Z}_1,\hat{Z}_2) &\equiv \lim_{\sigma\to\infty} 2\sigma^2
\mcK^{(2)}_{\text{Ai}}(\sigma\hat{Z}_1,\sigma\hat{Z}_2) \nn \\
 &= \frac{1}{\pi^{3/2}} \, \exp\left[ - \, \frac{\hat{Y}_1^2+\hat{Y}_2^2}{2} \right] \int_0^{\infty} ds \, \exp\left[ - \, \frac{(\hat{Z}_1+s)^2+(\hat{Z}_2+s)^2}{2} \right] \nn \\
 &= \frac{1}{2\pi} \, \exp\left[ - \, \frac{\hat{Y}_1^2+\hat{Y}_2^2}{2} - \frac{(\hat{Z}_1-\hat{Z}_2)^2}{4}\right] \erfc\left( \frac{\hat{Z}_1 + \hat{Z}_2}{2} \right).
\end{align}
The limiting density is then given by
\be
R^{(2)}_{1,\,\text{edge}}(\hat{Z})
=
\mcK^{(2)}_{\text{edge}}(\hat{Z},\hat{Z}^*)
=
\frac{1}{2\pi}\text{erfc}(\hat{X}).
\ee
This agrees with \cite{Kanzieper03, Mehta} at the edge of the spectrum on the real line in the limit of strong non-Hermiticity,
and is independent of $\hat{Y}$, as expected, since the real axis is not ``special'' for this ensemble (as it is for $\beta=1$ and $\beta=4$).

As a third check, we determine the limit of the kernel in the vicinity of the eigenvalue
with the largest real part, see \cite{Bender} for the original proof.
In particular, for the real part $X$ of the eigenvalues,
we must also introduce a ``shift'' of the origin to the right,
to track the typical $X$-coordinate (call it $X_{\text{max}}$) of the eigenvalue with largest $X$.
And for the $Y$-coordinate,
we have to introduce a particularly strong scaling with $\sigma$,
essentially to ``pull in'' (towards the real axis) the very, very small number of
outlying eigenvalues that have $X$ values in the vicinity of $X_{\text{max}}$.
As we will see,
prior to this ``pulling in'' process,
the eigenvalues with such high $X$ values are, in fact, so far apart in the $Y$-direction
that they are completely independent.

The precise limit that we determine is given by (see \cite{Bender})
\begin{align}
\label{Mbeta2}
M^{(2)}(z_1,z_2) &\equiv \lim_{\sigma\rightarrow\infty}
\sigma a(\sigma)b(\sigma) \,
\exp\left[ \frac{iX_1Y_1}{\sigma} + \frac{iY_1^3}{3\sigma^3} + \frac{iX_2Y_2}{\sigma} + \frac{iY_2^3}{3\sigma^3} \right] \,  \mcK_{\text{Ai}}^{(2)}(Z_1,Z_2;\sigma),
\end{align}
where $Z_j \equiv a(\sigma) x_j + c(\sigma) + i\sigma b(\sigma) y_j$ and $z_j = x_j + iy_j$,
and $a(\sigma)$, $b(\sigma)$ and $c(\sigma)$ are specified in eq.\ (\ref{abc_def}).
As is pointed out in \cite{Bender}, the exponential prefactors that have been introduced here
ensure that $M^{(2)}(z_1,z_2)$ itself exists,
but will not affect the correlation functions.
We can write $M^{(2)}(z_1,z_2)$ as
\begin{align}
M^{(2)}(z_1,z_2) &= \lim_{\sigma\rightarrow\infty}
\sigma a(\sigma)b(\sigma) \, \frac{1}{\sigma\sqrt{\pi}} \, \int_0^{\infty} dt \, h(Z_1,t,\sigma) h(Z_2,t,\sigma) \nn \\
 &= \frac{1}{\sqrt{\pi}} \lim_{\sigma\rightarrow\infty}
a(\sigma)b(\sigma)d(\sigma) \, \int_0^{\infty} du \, m(x_1,y_1,u,\sigma) m(x_2,y_2,u,\sigma),
\end{align}
where $h(Z,t,\sigma)$ and $m(x,y,u,\sigma)$ were defined in eqs.\ (\ref{h_def}) and (\ref{m_def}) respectively,
and $d(\sigma)$ in eq.\ (\ref{d_def}).
Since
\be
a(\sigma)b(\sigma)d(\sigma) = \frac{\sigma^{7/2}}{(6\log\sigma)^{3/4}},
\ee
and by substituting for $m(x,y,u,\sigma)$ from eq.\ (\ref{m_formula}), we have
\begin{align}
M^{(2)}(z_1,z_2) &= \frac{1}{\sqrt{\pi}} \lim_{\sigma\rightarrow\infty}
\frac{\sigma^{7/2}}{(6\log\sigma)^{3/4}} \, \left(\frac{(6\log\sigma)^{5/8}}{\sigma^{7/4}}\right)^2
\exp\left[ {} - \frac{x_1+x_2+y_1^2+y_2^2}{2} \right] \nn \\
 & \qquad \times \int_0^{\infty} du \, \exp\left[ {} - u^2 - \left( \sqrt{6\log\sigma} + \frac{i(y_1+y_2)\sigma^{3/2}}{(6\log\sigma)^{1/4}} \right)u \right] \nn \\
 &= \frac{1}{\sqrt{\pi}} \exp\left[ {} - \frac{x_1+x_2+y_1^2+y_2^2}{2} \right] \lim_{\sigma\rightarrow\infty}
 \sqrt{6\log\sigma} \int_0^{\infty} du \, \exp\left[ -u^2 - Ku \right] \nn \\
 &= \tfrac12 \exp\left[ {} - \frac{x_1+x_2+y_1^2+y_2^2}{2} \right] \lim_{\sigma\rightarrow\infty}
\sqrt{6\log\sigma} \, e^{K^2/4} \erfc\left(\frac{K}{2}\right)
\end{align}
where
\be
K \equiv \sqrt{6\log\sigma} + \frac{i(y_1+y_2)\sigma^{3/2}}{(6\log\sigma)^{1/4}}.
\ee
Finally, using the large-argument asymptotic of the complementary error function
\be\label{erf_asymp}
\erfc(z) \sim \frac{1}{z\sqrt{\pi}} \, e^{-z^2},
\ee
we immediately arrive at
\begin{align}
\label{Mbeta2_final}
M^{(2)}(z_1,z_2) &= \frac{1}{\sqrt{\pi}} \, \exp\left[ {} - \frac{x_1+x_2+y_1^2+y_2^2}{2} \right]
  \lim_{\sigma\rightarrow\infty}\left[1 + \frac{i(y_1+y_2)\sigma^{3/2}}{(6\log\sigma)^{3/4}}\right]^{-1} \nn \\
  &= \frac{1}{\sqrt{\pi}} \, \exp\left[ {} - \frac{x_1+x_2+y_1^2+y_2^2}{2} \right] \, \delta_{y_1,-y_2},
\end{align}
where $\delta_{uv}$ is the Kr\"onecker delta function from eq.\ (\ref{kron_delta_def}).
The kernel $M^{(2)}(z_1,z_2)$ is known as the Poisson kernel.

As indicated in \cite{Bender}, the presence of the delta function here implies that the eigenvalues are independent under this scaling.
Furthermore, the fact that the $x$- and $y$-dependent parts factorise
(i.e.\ no cross terms in the exponent)
means that the $y$-dependency can then be trivially integrated out.
This leaves the $x$-dependency being a simple exponential,
from which it is easily shown (see \cite{Bender}, for example, for a simple proof)
that the eigenvalue with largest real part
has the Gumbel probability distribution.
This is similar to the result that the eigenvalue with largest modulus
also obeys the Gumbel distribution, see \cite{Rider}.

Finally, we will show that it is possible to recover the weakly non-Hermitian kernel
in the bulk of the spectrum.
Specifically, we consider\footnote{Intuitively, the scaling parameter $w$ here can be considered roughly equivalent to $N^{1/3}$, where $N$ is the matrix size.}
\begin{align}
H^{(2)}(z_1,z_2; \sigma) &\equiv \lim_{w\rightarrow\infty} \frac{1}{w^2} \mcK_{\text{Ai}}^{(2)} \left( \frac{z_1}{w} - w^2, \frac{z_2}{w} - w^2; \frac{\sigma}{w} \right) \nn \\
 &= \lim_{w\rightarrow\infty} \frac{1}{w^2} \, \frac{w}{\sigma\sqrt{\pi}} \, \exp\left[ {} - \frac{y_1^2+y_2^2}{2\sigma^2} \right] \int_{1}^{-\infty} (-w^2 \, ds) \, \Aid\left( \frac{z_1}{w} - sw^2, \frac{\sigma}{w}\right)
 \Aid\left( \frac{z_2}{w} - sw^2, \frac{\sigma}{w}\right)
\end{align}
where we changed integration variable from $t$ to $s=1-t/w^2$.
Assuming now that the limit $w\rightarrow\infty$ and the integral commute,
we can replace the deformed Airy functions with their large-$w$ asymptotic limits, eq.\ (\ref{Aid_bulk_limit}).
These asymptotic limits are valid only for $s>0$.
However, the contribution to the integral from $s<0$ is small,
because the Airy functions are exponentially decreasing here,
rather than oscillatory.
Therefore, we can change the lower limit from $-\infty$ to 0, with negligible effect at large $w$:
\begin{align}
H^{(2)}(z_1,z_2; \sigma) &= \lim_{w\rightarrow\infty} \frac{w}{\sigma\sqrt{\pi}} \, \exp\left[ {} - \frac{y_1^2+y_2^2}{2\sigma^2} \right] \nn \\
 & \qquad \qquad \times
\int_0^1 ds \, \frac{\exp[-s\sigma^2]}{\pi\sqrt{s}\,w} \,
\sin \left( \tfrac23 s^{3/2} w^3 - \sqrt{s} \, z_1 + \frac{\pi}{4} \right)
\sin \left( \tfrac23 s^{3/2} w^3 - \sqrt{s} \, z_2 + \frac{\pi}{4} \right).
\end{align}
We now use the trigonometric relation
\be
\sin\left( A + \frac{\pi}{4} \right) \sin\left( B + \frac{\pi}{4} \right) =
\tfrac12 \big[ \cos(A-B) + \sin(A+B) \big]
\ee
to rewrite the integrand as the sum of two parts:
\begin{align}
H^{(2)}(z_1,z_2; \sigma) &= \frac{1}{2\sigma\pi^{3/2}} \, \exp\left[ {} - \frac{y_1^2+y_2^2}{2\sigma^2} \right] \nn \\
 & \qquad \qquad \times \lim_{w\rightarrow\infty}
\int_0^1 ds \, \frac{\exp[-s\sigma^2]}{\sqrt{s}} \,
\left\{
\cos\left(\sqrt{s}(z_1 - z_2)\right)
+\sin\left(\tfrac43 s^{3/2} w^3 - \sqrt{s}(z_1+z_2)\right)
\right\}.
\end{align}
However, at large $w$, the sine function here becomes very highly oscillatory as a function of $s$,
and so this part of the integral vanishes.
We are therefore left with
\begin{align}
H^{(2)}(z_1,z_2; \sigma) &= \frac{1}{2\sigma\pi^{3/2}} \, \exp\left[ {} - \frac{y_1^2+y_2^2}{2\sigma^2} \right] \,
\int_0^1 ds \, \frac{\exp[-s\sigma^2]}{\sqrt{s}} \,
\cos\left(\sqrt{s}(z_1 - z_2)\right)
\end{align}
which is the well-known interpolating sine kernel, see \cite{FKS2}.

\sect{The interpolating Airy kernel for $\beta=4$}\label{KAib4}

In this section we derive the microscopic edge scaling limit of the kernel $\mcK_N^{(4)}(z_1,z_2)$.
Because of the form of the Jacobian there are no real eigenvalues in this ensemble,
but we will still magnify the vicinity of the right- (or left-) most point
of the ellipse of support of the density along the real axis, located at $\pm(1+\tau)\sqrt{N}$.

The finite-$N$ solution of this ensemble with $\beta=4$ can be summarised as follows,
where we will follow \cite{Kanzieper_2002_02}.
On inserting the bivariate weight function eq.\ (\ref{beta_eq_4_F_def}) into eqs.\ (\ref{GN_def}) and (\ref{WN_def}) we obtain
\begin{align}
G_{2N}^{(4)}(z_1,z_2) &= (z_2-z_2^*) \,w^{(4)}(z_2)\,  \hat{\mcK}_{2N}^{(4)}(z_1,z_2^*), \label{beta_eq_4_G_2N_from_K} \\
W_{2N}^{(4)}(z_1,z_2) &= -(z_1-z_1^*)(z_2-z_2^*) \, w^{(4)}(z_1)w^{(4)}(z_2) \,  \hat{\mcK}_{2N}^{(4)}(z_1^*,z_2^*)
\nn \\
 & \qquad \qquad {}
- \sqrt{w^{(4)}(z_1)w^{(4)}(z_2)} \, (z_1 - z_2) \, \delta^{(2)}(z_1 - z_2^*) \ , \label{beta_eq_4_W_2N from K}
\end{align}
given in terms of the anti-symmetric pre-kernel (for even $N$)
\be
\label{beta_eq_4_prekernel}
\hat{\mcK}_N^{(4)}(z_1,z_2) =
\sum_{k=0}^{\frac{N}{2}-1} \frac{1}{r_k}\,\big( q_{2k+1}(z_1)q_{2k}(z_2) - q_{2k}(z_1)q_{2k+1}(z_2) \big)\ .
\ee

It is easy to see when applying rules for Pfaffians to the resulting eq.\ (\ref{Rkb4})
that the prefactors $(z-z^*)w^{(4)}(z)$ in front of the matrix-kernel elements above
can be distributed more symmetrically.
Let us consider (without loss of generality) the case where $\iim z_j \ge 0$ for all arguments.
This leads to
\be
R_{k,N}^{(4)}(z_1,\ldots,z_k) = \Pf_{i,j=1,\ldots,k}
\left[   (-2i)\sqrt{|y_i y_j| w^{(4)}(z_i)w^{(4)}(z_j)}
\left(
\begin{array}{ll}
\hat{\mcK}_{2N}^{(4)}(z_i,z_j) &\hat{\mcK}_{2N}^{(4)}(z_i,z_j^*)\\
\hat{\mcK}_{2N}^{(4)}(z_i^*,z_j) &\hat{\mcK}_{2N}^{(4)}(z_i^*,z_j^*)
\\
\end{array}
\right)\right]\ ,
\label{Rkb4alt}
\ee
where all the elements are given in terms of the same pre-kernel eq.\ (\ref{beta_eq_4_prekernel}), but with different arguments.
We have also dropped the contact terms here. For the density we thus obtain
\be
R_{1,N}^{(4)}(z_1)={\mcK}_{2N}^{(4)}(z_1,z_1^*)
\ee
with kernel
\be
{\mcK}_{2N}^{(4)}(z_1,z_2^*)\equiv (-2i)\sqrt{|y_1 y_2| w^{(4)}(z_1)w^{(4)}(z_2)} \,
\hat{\mcK}_{2N}^{(4)}(z_1,z_2^*)\ .
\label{K2N_def}
\ee

The polynomials $q_k(z)$ appearing in eq.\ (\ref{beta_eq_4_prekernel})
are skew-orthogonal with respect to
the following anti-symmetric scalar product,
\be
\langle q_{2k+1},q_{2j}\rangle\equiv
\int_{\mathbb  C} \dsq z\ (z^*-z)w^{(4)}(z)\Big(
q_{2k+1}(z)\ q_{2j}(z^*)-q_{2k+1}(z^*)\ q_{2j}(z)
\Big)=r_j\delta_{jk}
\label{skewOPdef}
\ee
and
\be
\langle q_{2k},q_{2j}\rangle=0=\langle q_{2k+1},q_{2j+1}\rangle,
\ee
and are given by
\begin{align}\label{beta_4_skOPs}
q_{2k}(z) &= 2^k \, k! \sum_{m=0}^k \frac{1}{(2m)!!} \, \left( \frac{\tau}{2} \right)^m  \, H_{2m} \left( \frac{z}{\sqrt{2\tau}} \right), \nn \\
q_{2k+1}(z) &= \left( \frac{\tau}{2} \right)^{k+1/2} H_{2k+1} \left( \frac{z}{\sqrt{2\tau}} \right),
\end{align}
with (squared) norms
\be\label{beta_4_norm}
r_k = 2\pi (1-\tau)^{3/2}(1+\tau)^{1/2} \, (2k+1)! \ .
\ee
Inserting the skew-orthogonal polynomials into the pre-kernel eq.\ (\ref{beta_eq_4_prekernel}) gives a double sum, which in general cannot be reduced to a single sum.
Note that the definition of such skew-orthogonal polynomials is not unique  for $\beta=4$ and $\beta=1$ \cite{Kanzieper_2002_02,AKP}. In our choice they have parity according to their degree, $q_k(-z)=(-)^kq_k(z)$.

As a first step we will rewrite the kernel eq.\ (\ref{K2N_def})
in order to facilitate the asymptotic analysis.
The asymptotic limits of the remaining matrix elements then easily follow.
For even $N$, we have
\begin{align}
{\mcK}_N^{(4)}(z_1,z_2) &=  \frac{(-2i) \sqrt{|y_1 y_2| w^{(4)}(z_1) w^{(4)}(z_2)} }{2\pi(1-\tau)^{3/2}(1+\tau)^{1/2}} \nn \\
 &\times \sum_{\substack{m=0 \\ m\text{ even}}}^{N-2} \sum_{\substack{k=m+1 \\ k \text{ odd}}}^{N-1} \left\{ \frac{1}{k!!} \left(\frac{\tau}{2}\right)^{k/2} H_{k}\left( \frac{z_1}{\sqrt{2\tau}} \right)  \frac{1}{m!!} \left(\frac{\tau}{2}\right)^{m/2} H_{m}\left( \frac{z_2}{\sqrt{2\tau}} \right) - (z_1 \leftrightarrow z_2) \right\},
 \label{b4_kernel_explicit}
\end{align}
where we swapped the order of the double summation over $k$ and $m$ (which runs over a triangular region) and relabelled.

We now apply a procedure similar to that for the $\beta=2$ case in the previous section, and so will give only an outline. We switch from integer summation indices $m$ and $k$ to indices $s$ and $t$ that are (strictly positive) integer multiples of $\Delta s = \Delta t = 2/N^{1/3}$:
\begin{align}
m &= N - sN^{1/3} && \text{for} && s = \Delta s, 2\Delta s, \ldots, \tfrac12 N \Delta s = N^{2/3},\nn \\
k &= N + 1 - tN^{1/3} && \text{for} && t = \Delta t, 2\Delta t, \ldots, s.
\end{align}
By redefining $m\equiv m(s,N)$ and $k\equiv k(t,N)$ for $0 \le s \le N^{2/3}$ and $0 \le t \le s$ as
\begin{align}
m &= \inteven(N - sN^{1/3}),\nn \\
k &= \intodd(N + 1 - tN^{1/3}),
\end{align}
where the functions $\inteven(x)$ and $\intodd(x)$ give the largest even and odd integers less than or equal to $x$,
we can write the double sum as a two-dimensional integral over a step function
\begin{align}
{\mcK}_N^{(4)}(z_1,z_2) &=  \frac{N^{2/3}}{4} \, \frac{(-2i) \sqrt{|y_1 y_2| w^{(4)}(z_1) w^{(4)}(z_2) }}{2\pi(1-\tau)^{3/2}(1+\tau)^{1/2}} \nn \\
 &\times \int_{0}^{N^{2/3}} ds \int_{0}^{s} dt
   \left\{ \frac{1}{k!!} \left(\frac{\tau}{2}\right)^{k/2} H_{k}\left( \frac{z_1}{\sqrt{2\tau}} \right)  \frac{1}{m!!} \left(\frac{\tau}{2}\right)^{m/2} H_{m}\left( \frac{z_2}{\sqrt{2\tau}} \right) - (z_1 \leftrightarrow z_2) \right\},
\end{align}
which is still exact.

The large-$N$ limit is defined as
\be
\mcK^{(4)}_{\text{Ai}}(Z_1,Z_2) \equiv \lim_{\substack{N\to\infty \\ \tau\to1}} \frac{1}{N^{1/3}} e^{-iN^{1/3}(Y_1+Y_2)} \mcK_N^{(4)}\left(
(1+\tau) \sqrt{N} + \frac{Z_1}{N^{1/6}},
(1+\tau) \sqrt{N} + \frac{Z_2}{N^{1/6}}\right),
\ee
where, as before, we have shifted the kernel by an oscillatory phase before taking the large-$N$ limit.
Once again all these phases cancel, after extracting them from the rows and columns of the Pfaffian in eq.\ (\ref{Rkb4alt}).
At large $N$, we have for the prefactors
\be
\frac{1}{(1-\tau)^{3/2}(1+\tau)^{1/2}} \sim \frac{N^{1/2}}{\sqrt{2}\,\sigma^3}
\ee
and
\be
|y_1 y_2| = \frac{|Y_1 Y_2|}{N^{1/3}}.
\ee
Again we can take the limit inside the integral, and so must consider the
large-$N$ limits of the individual factors in the integrand.
At fixed $s$ and $t$ and at large $N$, we have for even $m$
\be\label{double_fac_asymp_even}
\frac{1}{m!!}
\sim \left( \frac{2}{\pi} \right)^{1/4} \frac{1}{m^{1/4}\sqrt{m!}}
\sim\left( \frac{2}{\pi} \right)^{1/4} \frac{1}{N^{1/3}} \, \frac{m^{1/12}}{\sqrt{m!}}\ ,
\ee
and for odd $k$
\be\label{double_fac_asymp_odd}
\frac{1}{k!!}
\sim \left(\frac{\pi}{2}\right)^{1/4} \, \frac{1}{k^{1/4}\sqrt{k!}}
\sim\left(\frac{\pi}{2}\right)^{1/4} \,  \frac{1}{N^{1/3}} \, \frac{k^{1/12}}{\sqrt{k!}}\ ,
\ee
where the first step in eq.\ (\ref{double_fac_asymp_even}) follows by applying
Stirling's formula $n! \sim \sqrt{2\pi n}(n/e)^n$ twice, and eq.\ (\ref{double_fac_asymp_odd})
then follows from eq.\ (\ref{double_fac_asymp_even}).
We can therefore write
\begin{align}
& {\mcK}_{\text{Ai}}^{(4)}(Z_1,Z_2) =
\lim_{\substack{N\to\infty \\ \tau\to1}}
\left( \frac{1}{N^{1/3}} \, \frac{N^{2/3}}{4} (-2i) \sqrt{\frac{|Y_1 Y_2|}{N^{1/3}}} \, \frac{1}{2\pi}
\, \frac{N^{1/2}}{\sqrt{2}\,\sigma^3} \, \frac{\sqrt{2\pi}}{N^{2/3}} \right) \nn \\
 & \qquad \times
 \left\{
\int_0^{\infty} ds \int_0^s dt
\lim_{\substack{N\rightarrow\infty \\ \tau\to1}}
\Big( e^{-iN^{1/3}Y_1} h_k(z_1) \Theta(k) \Big)
\lim_{\substack{N\rightarrow\infty \\ \tau\to1}}
\Big( e^{-iN^{1/3}Y_2} h_m(z_2) \Theta(m) \Big)
- (z_1 \leftrightarrow z_2)
\right\}
\end{align}
where $h_j(z)$ was defined in eq.\ (\ref{beta_eq_2_hj_def}).
Applying our result eq.\ (\ref{hjlim}) from the previous section we finally arrive at
\begin{align}
\label{beta_eq_4_K_final}
\mcK^{(4)}_{\text{Ai}}(Z_1,Z_2) &= \frac{-i\sqrt{|Y_1Y_2|}}{4\sqrt{\pi}\,\sigma^3} \, \exp\left[ - \, \frac{Y_1^2+Y_2^2}{2\sigma^2} + \frac{\sigma^6}{6} + \frac{\sigma^2(Z_1+Z_2)}{2} \right] \nn \\
 & \qquad \times \left\{ \int_0^{\infty} ds \, \e^{\sigma^2 s/2} \, \Ai\left( Z_2 + \frac{\sigma^4}{4} + s \right)\,\int_0^{s} dt \, \e^{\sigma^2 t/2} \Ai\left( Z_1 + \frac{\sigma^4}{4} + t \right) - (Z_1 \leftrightarrow Z_2) \right\}.
\end{align}
The limits for the remaining matrix-kernel elements in eq.\ (\ref{Rkb4alt}) are obtained simply by complex conjugation of the arguments\footnote{The microscopic limit can also be taken including the contact terms in eq.\ (\ref{Rkb4}), observing that
$N^{-1/3}\delta^{(2)}(z_1-z_2^*)=N^{-1/3}\delta(N^{-1/6}(X_1-X_2))\delta(N^{-1/6}(Y_1+Y_2))=
\delta^{(2)}(Z_1-Z_2^*)$
with the scaling eq.\ (\ref{edgelim}).}.

We note that the same limiting kernel is obtained in eq.\ (\ref{beta_eq_4_K_final}) when taking the microscopic limit at negative arguments around $-(1+\tau) \sqrt{N}$. This is due to the fact that our skew-orthogonal polynomials have parity, and hence the pre-kernel in eq.\ (\ref{beta_eq_4_prekernel}) changes sign, $\hat{\mcK}_N^{(4)}(-z_1,-z_2)
=-\hat{\mcK}_N^{(4)}(z_1,z_2)$.
The overall minus sign in front of the matrix-kernel is a phase that
can be defined away, once one maps back to eigenvalues in the upper half of the complex plane.

Let us give as an example the limiting microscopic density at the edge of the spectrum:
\begin{align}
R_{1,\text{Ai}}^{(4)}(Z) &=  \mcK_{\text{Ai}}^{(4)}(Z,Z^*)\nn\\
 &= \frac{Y}{2\sqrt{\pi}\,\sigma^3} \, \exp\left[ - \, \frac{Y^2}{\sigma^2} + \frac{\sigma^6}{6} + \sigma^2 X \right] \nn \\
 & \qquad \times \iim \left[ \int_0^{\infty} ds \, \e^{\sigma^2 s/2} \, \Ai\left( Z^* + \frac{\sigma^4}{4} + s \right)\,\int_0^{s} dt \, \e^{\sigma^2 t/2} \Ai\left( Z + \frac{\sigma^4}{4} + t \right)\right]. \label{GinSE_weak_density}
\end{align}
By dropping the modulus sign around the leading $Y$, this expression becomes valid in the whole of the
complex plane, and not just in the upper half-plane.
As a check we will now show that this reduces to the known $\beta=4$ Airy density in the Hermitian limit ${\sigma\to0}$.
The Hermitian limit of the full matrix-kernel is relegated to Appendix \ref{Hermlimb4}, being more technically involved.


\subsection{Hermitian and strongly non-Hermitian limits}
\label{beta_eq_4_density_Hermitian}

Let us consider the Hermitian limit of the eigenvalue density eq.\ (\ref{GinSE_weak_density}).
First, consider some fixed $Z$ away from the real axis, i.e.\ $|Y|>0$. As $\sigma\rightarrow 0$, the term in square brackets inside the imaginary part in eq.\ (\ref{GinSE_weak_density}) will tend to something finite. The $Y$-dependent part of the exponential prefactor will tend to zero, since $|Y|>0$. Hence, the density of complex (i.e.\ non-real) eigenvalues tends to zero, as expected.

Now let us try to understand what happens close to, and on, the real axis.  Of course, for any $\sigma>0$, the eigenvalue density on the real axis will remain zero.
In fact, for small but non-zero $\sigma$, the density takes the form of two ridges, parallel to the real axis.
So na\"ively setting $Y=0$ and then taking the limit $\sigma\rightarrow 0$ will not generate the density of real eigenvalues when $\sigma=0$.
In this sense the Hermitian limit on the real axis is discontinuous.

Instead, let us assume that both $\sigma$ and $Y$ are small and of the same order, $\sigma\sim Y\ll 1$, but both are non-zero. We can then approximate the density in eq.\ (\ref{GinSE_weak_density}), keeping only leading order terms.
With a slight abuse of notation, we write
\begin{align}
\lim_{\sigma,Y\ll 1} R_{1,\text{Ai}}^{(4)}(Z) &\approx \frac{Y}{2\sqrt{\pi}\,\sigma^3} \, \exp\left[ - \, \frac{Y^2}{\sigma^2} \right] \iim \left[ \int_0^{\infty} ds \, \Ai( X - iY + s )\,\int_0^{s} dt \, \Ai( X + iY + t )\right]\nn\\
&\approx \frac{Y^2}{2\sqrt{\pi}\,\sigma^3} \,
\exp\left[- \, \frac{Y^2}{\sigma^2}\right]
\int_0^{\infty} ds  \int_0^{s} dt \, \Big( \Ai(X+s)\Ai\hspace{0mm}'(X+t) - \Ai\hspace{0mm}'(X+s)  \Ai(X+t) \Big).\nn\\
\end{align}
In the second step we used a Taylor series expansion of the Airy functions about $X+s$ and $X+t$.
Note that the expression for the density has now factorised into separate $X$- and $Y$-dependent parts.  For the first term inside the double integral, we can do the $t$-integral easily,
$\int_0^{s} dt \, \Ai\hspace{0mm}'(X+t) = \Ai(X+s) - \Ai(X)$\ .
The second term can be integrated by parts in the $s$-variable
\be
\label{AiryInt}
- \int_0^{\infty} ds \, \Ai\hspace{0mm}'(X+s) \, f(s)
= - \Big[ f(s) \, \Ai(X+s) \Big]_{s=0}^{s=\infty} + \int_0^{\infty} ds \, \Ai(X+s) \, f'(s)
=\int_0^{\infty} ds \, [\Ai(X+s)]^2\ ,
\ee
where for compactness we defined
$f(s) \equiv \int_0^{s} dt \, \Ai(X+t)$.
Assembling these results gives
\be
\lim_{\sigma,Y\ll 1}R_{1,\text{Ai}}^{(4)}(Z) \approx
\frac{Y^2}{2\sqrt{\pi}\,\sigma^3} \, \exp\left[ - \, \frac{Y^2}{\sigma^2} \right] \left\{ 2 \int_0^{\infty} ds \, [\Ai(X+s)]^2 - \Ai(X) \int_0^{\infty} ds \, \Ai(X+s) \right\}.
\ee
It is easy to see that the prefactor (giving two peaks at $Y = \pm \sigma$)
results in a \textit{single} delta function on the real axis as eq.\ (\ref{deltadef1}) in the limit $\sigma\rightarrow 0$, i.e.
\be
\label{deltadef2}
\lim_{\sigma\rightarrow 0} \frac{2Y^2}{\sqrt{\pi}\,\sigma^3} \, \exp\left[ - \, \frac{Y^2}{\sigma^2} \right] = \delta(Y).
\ee
Hence the Hermitian limit of the eigenvalue density eq.\ (\ref{GinSE_weak_density}) is given by
\be
\lim_{\sigma\to0}R_{1,\text{Ai}}^{(4)}(Z) = \left\{ \frac{1}{2} \int_0^{\infty} ds \, [\Ai(X+s)]^2 - \frac{1}{4} \Ai(X) \int_X^{\infty} ds \, \Ai(s) \right\} \delta(Y) \
\ee
which agrees with the known density for $\beta=4$ \cite{Anderson} multiplied by a delta function, where the first term is, in fact, precisely one half of the limiting density for the $\beta=2$ ensemble, see eq.\ (\ref{microb2real}).

For the strongly non-Hermitian limit, we find for the kernel
\begin{align}
& \mcK_{\text{edge}}^{(4)}(\hat{Z}_1,\hat{Z}_2) \equiv \lim_{\sigma\rightarrow\infty} 2 \sigma^2 \mcK_{\text{Ai}}^{(4)}(\sigma\hat{Z}_1,\sigma\hat{Z}_2) \nn \\
 &= \frac{-i\sqrt{|\hat{Y}_1 \hat{Y}_2}|}{4\pi^{3/2}} \exp\left[ - \, \frac{\hat{Y}_1^2+\hat{Y}_2^2}{2} \right] \left\{ \int_0^{\infty} du \, \exp\left[ - \, \frac{(u+\hat{Z}_2)^2}{2} \right] \int_0^u dv \, \exp\left[ - \, \frac{(v+\hat{Z}_1)^2}{2} \right] - (\hat{Z}_1 \leftrightarrow \hat{Z}_2) \right\} \nn \\
 &= \frac{-i\sqrt{|\hat{Y}_1 \hat{Y}_2|}}{4\sqrt{2}\pi} \exp\left[ - \, \frac{\hat{Y}_1^2+\hat{Y}_2^2}{2} \right] \left\{ \int_0^{\infty} du \, \exp\left[ - \, \frac{(u+\hat{Z}_1)^2}{2} \right] \erfc\left( \frac{u+\hat{Z}_2}{\sqrt{2}} \right) - (\hat{Z}_1 \leftrightarrow \hat{Z}_2) \right\}
\end{align}
and the corresponding spectral density
\be
R^{(4)}_{1,\,\text{edge}}(\hat{Z}) = \frac{\hat{Y}}{2\sqrt{2}\pi} \exp\left[ - \hat{Y}^2 \right] \iim\left\{ \int_0^{\infty} du \, \exp\left[ - \, \frac{(u+\hat{Z})^2}{2} \right]  \erfc\left( \frac{u+\hat{Z}^*}{\sqrt{2}} \right)  \right\}.
\ee
Unlike for the $\beta=2$ case, in this ensemble the real axis is ``special'',
and so we still see a repulsion of eigenvalues from the real axis even in this strongly non-Hermitian limit.
If we consider the behaviour of this result at large $\hat{Y}$ by using the fact that, at large $y \equiv \iim z$,
\be
\erf(z) \sim \frac{i}{\sqrt{\pi} \, y} e^{-z^2} \qquad\text{and}\qquad \erfc(z) \sim -\erf(z),
\ee
we can then show that the kernel behaves as
\begin{align}
\mcK_{\text{edge}}^{(4)}(\hat{Z}_1,\hat{Z}_2) &\sim \frac{\hat{Y}_1 - \hat{Y}_2}{8\pi\sqrt{|\hat{Y}_1 \hat{Y}_2|}} \, \exp \left[ - \, \frac{\hat{Y}_1^2 + \hat{Y}_2^2}{2}  - \frac{(\hat{Z}_1 - \hat{Z}_2)^2}{4} \right] \erfc\left( \frac{\hat{Z}_1+\hat{Z}_2}{2} \right) \nn \\
 &=  \frac{\hat{Y}_1 - \hat{Y}_2}{4\sqrt{|\hat{Y}_1 \hat{Y}_2|}} \, \mcK_{\text{edge}}^{(2)}(\hat{Z}_1,\hat{Z}_2),
\end{align}
and the density as
\be
R^{(4)}_{1,\,\text{edge}}(\hat{Z}) \sim \frac{1}{4\pi} \, \erfc(\hat{X}).
\ee
We have not seen this result in the existing literature.
It is essentially the same as the result for $\beta=2$,
demonstrating the universality of this result at the edge of the spectrum in the complex plane,
far from the real axis.

Finally we consider the alternative $\sigma\rightarrow\infty$ limit, where we
shift and scale the eigenvalues to remain within the vicinity of the
eigenvalue with the largest real part.
Following eq.\ (\ref{Mbeta2}) we have
\begin{align}
M^{(4)}(z_1,z_2) &\equiv \lim_{\sigma\rightarrow\infty} \sigma \, a(\sigma) b(\sigma) \exp\left[ \frac{iX_1Y_1}{\sigma} + \frac{iY_1^3}{3\sigma^3} + \frac{iX_2Y_2}{\sigma} + \frac{iY_2^3}{3\sigma^3} \right] \,  \mcK_{\text{Ai}}^{(4)}(Z_1,Z_2;\sigma) \nn \\
 &= \frac{-i\sqrt{|y_1 y_2|}}{4\sqrt{\pi}} \, \exp\left[ {} - \frac{x_1+x_2+y_1^2+y_2^2}{2} \right] \lim_{\sigma\rightarrow\infty} \sigma^{3/2} (6\log\sigma)^{1/4} \nn \\
 & \qquad \times \left\{ \int_0^{\infty} du \, \exp\left[ {} - \frac{u^2}{2} - K_2 u \right] \int_0^{u} dv \, \exp\left[ {} - \frac{v^2}{2} - K_1 v \right]  - (z_1 \leftrightarrow z_2) \right\},
\end{align}
where
\begin{align}
K_1 \equiv \frac{\sqrt{6\log\sigma}}{2} + \frac{iy_1\sigma^{3/2}}{(6\log\sigma)^{1/4}}, \\
K_2 \equiv \frac{\sqrt{6\log\sigma}}{2} + \frac{iy_2\sigma^{3/2}}{(6\log\sigma)^{1/4}}.
\end{align}
The inner integral over $v$ can be done exactly,
giving the difference of two error functions
(or, equivalently, the difference of two complementary error functions).
We then use the large-argument asymptotic behaviour of these, eq.\ (\ref{erf_asymp}),
which allows us to do the outer integral over $u$.
This leads to more complementary error functions,
which we again replace with their asymptotic limits.
The result is then
\be
M^{(4)}(z_1,z_2) = \frac{-i\sqrt{|y_1 y_2|}}{4\sqrt{\pi}} \, \exp\left[ {} - \frac{x_1+x_2+y_1^2+y_2^2}{2} \right] \lim_{\sigma\rightarrow\infty} \sigma^{3/2} (6\log\sigma)^{1/4} \, \frac{K_1-K_2}{K_1 K_2 (K_1+K_2)}.
\ee
Now,
\be
\frac{K_1-K_2}{K_1 K_2 (K_1+K_2)} \sim
\begin{cases}
\displaystyle {} - \frac{\sqrt{6\log\sigma}(y_1 - y_2)}{\sigma^3 y_1 y_2 (y_1 + y_2)} & \text{if }y_1 \ne  -y_2, \\ \\
\displaystyle \frac{2i}{(6\log\sigma)^{1/4}\sigma^{3/2}y_1} & \text{if } y_1 = -y_2,
\end{cases}
\ee
from which it immediately follows that
\be
M^{(4)}(z_1,z_2) = \frac{1}{2\sqrt{\pi}} \, \exp\left[ {} - \frac{x_1+x_2+y_1^2+y_2^2}{2} \right] \, \delta_{y_1,-y_2} = \tfrac12 \, M^{(2)}(z_1,z_2).
\ee
This is essentially the same result as for $\beta=2$,
which would have been expected from universality arguments,
and so the eigenvalue with largest real part also obeys the Gumbel distribution.
We believe that this is a new result, although, as for the $\beta=2$ ensemble discussed earlier,
it is already known, see \cite{Rider}, that the eigenvalue with largest modulus
also obeys the Gumbel distribution for this ensemble.

\sect{The interpolating Airy kernel for $\beta=1$}\label{KAib1}

For $\beta=1$, we have the pre-kernel (recall that we consider only even matrix size $N$ in this paper)
\be\label{beta_eq_1_K_def}
\hat{\mcK}^{(1)}_N(z_1,z_2) = \sum_{k=0}^{\frac{N}{2}-1} \frac{1}{r_k}\,\big( q_{2k+1}(z_1)q_{2k}(z_2) - q_{2k}(z_1)q_{2k+1}(z_2) \big).
\ee
On inserting the $\beta=1$ weight function eq.\ (\ref{beta_eq_1_F_def}) into eqs.\ (\ref{GN_def}) and (\ref{WN_def}), we see that the other two elements of the matrix-kernel, $G_N^{(1)}(z_1,z_2)$ and $W_N^{(1)}(z_1,z_2)$, each split into separate terms, which we write as follows:
\be\label{G_split_R_C}
G_N^{(1)}(z_1,z_2) = G_N^{\RR\,(1)}(z_1,x_2) \, \delta(y_2) + G_N^{\CC\,(1)}(z_1,z_2)
\ee
where
\begin{align}
G_N^{\RR\,(1)}(z_1,x_2) &\equiv -  w^{(1)}(x_2) \intR dt \, \sgn(x_2 - t) w^{(1)}(t) \hat{\mcK}_N^{(1)}(z_1,t), \label{GNR_expression} \\
G_N^{\CC\,(1)}(z_1,z_2) &\equiv 2i \sgn(y_2) w^{(1)}(z_2)w^{(1)}(z_2^*) \hat{\mcK}_N^{(1)}(z_1,z_2^*) \label{GNC_expression},
\end{align}
and
\begin{align}\label{W_split_R_C}
W_N^{(1)}(z_1,z_2) &= W_N^{\RR\RR\,(1)}(x_1,x_2) \, \delta(y_1)\delta(y_2) + W_N^{\RR\CC\,(1)}(x_1,z_2) \, \delta(y_1) \nn \\
 & \qquad {} - W_N^{\RR\CC\,(1)}(x_2,z_1) \, \delta(y_2) + W_N^{\CC\CC\,(1)}(z_1,z_2) - \mcF^{(1)}(z_1,z_2)
\end{align}
where
\begin{align}
W_N^{\RR\RR\,(1)}(x_1,x_2) &\equiv w^{(1)}(x_1) \int_{\RR} dt \, \sgn(x_1-t) w^{(1)}(t) G_N^{\RR\,(1)}(t,x_2), \label{WNRR_expression} \\
W_N^{\RR\CC\,(1)}(x_1,z_2) &\equiv -2i \sgn(y_2) w^{(1)}(z_2) w^{(1)}(z_2^*) G_N^{\RR\,(1)}(z_2^*,x_1), \label{WNRC_expression} \\
W_N^{\CC\CC\,(1)}(z_1,z_2) &\equiv 4 \sgn(y_1) \sgn(y_2) w^{(1)}(z_1) w^{(1)}(z_1^*) w^{(1)}(z_2) w^{(1)}(z_2^*) \hat{\mcK}_N^{(1)}(z_1^*,z_2^*). \label{WNCC_expression}
\end{align}
The $k$-point correlation functions $R_{k,N}^{(1)}(z_1,\ldots,z_k)$ are given by eq.\ (\ref{Rkb4}); in particular, the eigenvalue density is given by (minus) the function $G_N^{(1)}(z_1,z_2)$ evaluated at equal arguments:
\be\label{b1_density_from_GN_generic}
R_{1,N}^{(1)}(z) = -G_N^{(1)}(z,z) = - \left[ G_N^{\RR\,(1)}(x,x) \delta(y) + G_N^{\CC\,(1)}(z,z) \right],
\ee
indicating that, on average, a non-zero fraction of the eigenvalues are real.  The polynomials that appear in eq.\ (\ref{beta_eq_1_K_def}) are skew-orthogonal in the complex plane with respect to the weight function $\mcF^{(1)}(z_1,z_2)$ in eq.\ (\ref{beta_eq_1_F_def})
\be
\langle q_{2k+1}, q_{2j} \rangle \equiv \int_{\CC^2} \dsq z_1 \, \dsq z_2 \, \mcF^{(1)}(z_1,z_2) \, q_{2k+1}(z_1) q_{2j}(z_2) = r_k \delta_{jk},
\ee
\be
\langle q_{2k}, q_{2j} \rangle = \langle q_{2k+1}, q_{2j+1} \rangle = 0,
\ee
and are given by \cite{Forrester08}
\begin{align}\label{beta_1_skOPs}
q_{2k}(z) &= \left( \frac{\tau}{2} \right)^k H_{2k} \left( \frac{z}{\sqrt{2\tau}} \right), \nn \\
q_{2k+1}(z) &= \left( \frac{\tau}{2} \right)^{k+1/2} H_{2k+1} \left( \frac{z}{\sqrt{2\tau}} \right) - 2k \left( \frac{\tau}{2} \right)^{k-1/2} H_{2k-1} \left( \frac{z}{\sqrt{2\tau}} \right)
\end{align}
with (squared) norms
\be\label{beta_1_norm}
r_k = (2k)! \, 2\sqrt{2\pi} (1+\tau).
\ee

\subsection{Relationships between $\beta=1$ and $\beta=2$ kernels at finite $N$}

In this section, we derive some useful relationships between the finite-$N$ kernels
of the $\beta=1$ and $\beta=2$ ensembles.
The principal results here are eqs.\ (\ref{prekernel_exact_compact}) and (\ref{GN_from_KN_finite_N})
which are used later to write some results for $\beta=1$ in terms of results for $\beta=2$ that we already derived.

First we derive an exact relationship between the pre-kernel for $\beta=2$ (similar to the kernel eq.\ (\ref{KNb2}) but without the weight function factors)
\be\label{prekernel_beta_eq_2_exact}
\hat{\mcK}_N^{(2)}(z_1,z_2) \equiv \sum_{j=0}^{N-1} \frac{1}{c_j} \, p_j(z_1) p_j(z_2) = \frac{1}{\pi(1-\tau^2)^{1/2}} \sum_{j=0}^{N-1} \frac{1}{j!} \, \left( \frac{\tau}{2} \right)^j H_j\left( \frac{z_1}{\sqrt{2\tau}} \right) H_j\left( \frac{z_2}{\sqrt{2\tau}} \right)
\ee
and that for $\beta=1$ (for even $N$ only) which, from eqs.\ (\ref{beta_eq_1_K_def}), (\ref{beta_1_skOPs}) and (\ref{beta_1_norm}), is given by
\be\label{prekernel_beta_eq_1_exact}
\hat{\mcK}_N^{(1)}(z_1,z_2) = \frac{1}{2\sqrt{2\pi}(1+\tau)} \sum_{j=0}^{N-2} \frac{\left( \frac{\tau}{2} \right)^{j+1/2}}{j!} \,  \left\{ H_{j+1}\left( \frac{z_1}{\sqrt{2\tau}} \right) H_j\left( \frac{z_2}{\sqrt{2\tau}} \right) - H_j\left( \frac{z_1}{\sqrt{2\tau}} \right) H_{j+1}\left( \frac{z_2}{\sqrt{2\tau}} \right) \right\}.
\ee
Using the recurrence relation for Hermite polynomials (eq.\ 8.952.2 of \cite{Grad})
\be\label{Hermite_RR}
H_{n+1}(z) = 2zH_n(z) - 2nH_{n-1}(z),
\ee
we replace both occurences of $H_{j+1}$ on the right-hand side of eq.\ (\ref{prekernel_beta_eq_1_exact}) with Hermite polynomials of lower degree. After multiplying both sides by $2\sqrt{2\pi}(1+\tau)$, this then gives
\begin{align}
2\sqrt{2\pi}(1+\tau) \, \hat{\mcK}_N^{(1)}(z_1,z_2) &= \pi \sqrt{1-\tau^2} (z_1-z_2) \hat{\mcK}_{N-1}^{(2)}(z_1,z_2) \nn \\
 & \qquad {} - \tau \sum_{j=1}^{N-2} \frac{1}{(j-1)!} \left( \frac{\tau}{2} \right)^{j-1/2} \left\{ H_{j-1}\left( \frac{z_1}{\sqrt{2\tau}} \right) H_j\left( \frac{z_2}{\sqrt{2\tau}} \right) - (z_1 \leftrightarrow z_2) \right\} \nn \\
 &= \pi \sqrt{1-\tau^2} (z_1-z_2) \hat{\mcK}_{N-1}^{(2)}(z_1,z_2) + 2 \sqrt{2\pi} \, \tau \, (1+\tau) \, \hat{\mcK}_N^{(1)}(z_1,z_2) \nn \\
 & \qquad - \frac{\tau}{(N-2)!} \left( \frac{\tau}{2} \right)^{N-3/2} \left\{ H_{N-1}\left( \frac{z_1}{\sqrt{2\tau}} \right) H_{N-2}\left( \frac{z_2}{\sqrt{2\tau}} \right) - (z_1 \leftrightarrow z_2) \right\},
\end{align}
where we identified suitable combinations of terms with $\hat{\mcK}_{N-1}^{(2)}(z_1,z_2)$ and $\hat{\mcK}_N^{(1)}(z_1,z_2)$. Then we simply bring both occurences of $\hat{\mcK}_N^{(1)}(z_1,z_2)$ to the left-hand side, to give
\begin{align}\label{prekernel_exact_compact}
2\sqrt{2\pi}(1-\tau^2)\hat{\mcK}_N^{(1)}(z_1,z_2) &= \pi \sqrt{1-\tau^2} \, (z_1 - z_2) \, \hat{\mcK}_{N-1}^{(2)}(z_1,z_2) \nn \\
 & \qquad {} - \frac{2}{(N-2)!} \left( \frac{\tau}{2} \right)^{N-1/2} \left\{ H_{N-1}\left( \frac{z_1}{\sqrt{2\tau}} \right)H_{N-2}\left( \frac{z_2}{\sqrt{2\tau}} \right) - (z_1 \leftrightarrow z_2) \right\}.
\end{align}
Whilst eq.\ (\ref{prekernel_exact_compact}) has a relatively simple form, we note that we can use the Christoffel-Darboux formula for Hermite polynomials
\be
\sum_{j=0}^{N} \frac{1}{j!\,2^j} \, H_j(u) H_j(v) = \frac{1}{N!\,2^{N+1}} \, \frac{H_N(v) H_{N+1}(u) - H_N(u) H_{N+1}(v)}{u-v}
\ee
to replace the term in curly brackets in eq.\ (\ref{prekernel_exact_compact}) with a sum, as follows:
\begin{align}\label{prekernel_exact}
2\sqrt{2\pi}(1-\tau^2)\hat{\mcK}_N^{(1)}(z_1,z_2) &= \pi \sqrt{1-\tau^2} \, (z_1 - z_2) \, \hat{\mcK}_{N-1}^{(2)}(z_1,z_2) \nn \\
 & \qquad {} - \tau^{N-1} (z_1 - z_2) \sum_{j=0}^{N-2} \frac{1}{j!\,2^j} \, H_j \left( \frac{z_1}{\sqrt{2\tau}} \right)  H_j \left( \frac{z_2}{\sqrt{2\tau}} \right).
\end{align}
Although this step appears to be a retrograde one, the form eq.\ (\ref{prekernel_exact}) of the relationship will be the most useful when considering the densities of complex eigenvalues. Note that, for $\tau<1$, we have $\tau^{N-1} < \tau^{j}$ since $j<N-1$, and so the second term in eq.\ (\ref{prekernel_exact}) (involving the sum) is smaller than the first (involving $\hat{\mcK}_{N-1}^{(2)}$ which also contains a sum, see eq.\ (\ref{prekernel_beta_eq_2_exact})).

Second, we derive a relationship between the $\beta=2$ kernel in eq.\ (\ref{KNb2}) and the $\beta=1$ function $G_N^{\RR\,(1)}(z_1,x_2)$ in eq.\ (\ref{GNR_expression}).  Both of these include the weight functions (unlike the pre-kernels), and the latter also involves the sign function and an integral, and so the result eq.\ (\ref{GN_from_KN_finite_N}) that we will derive is not the same as eq.\ (\ref{prekernel_exact}), although they do have similar forms.  The relation that we derive is required for real $x_2$ only. We begin by inserting the $\beta=1$ pre-kernel eq.\ (\ref{prekernel_beta_eq_1_exact}) into eq.\ (\ref{GNR_expression}), giving
\be\label{GN_explicit}
- \, \frac{G_N^{\RR\,(1)}(z_1,x_2)}{w^{(1)}(x_2)} = \frac{1}{2\sqrt{2\pi}(1+\tau)} \sum_{j=0}^{N-2} \frac{1}{j!} \, \left( \frac{\tau}{2} \right)^{j+1/2} \left\{ H_{j+1}\left( \frac{z_1}{\sqrt{2\tau}} \right) I_j(x_2;\tau) - H_{j}\left( \frac{z_1}{\sqrt{2\tau}} \right) I_{j+1}(x_2;\tau) \right\},
\ee
where we have defined the function $I_j(x;\tau)$ for real $x$ as
\be\label{Ij_def}
I_j(x;\tau) \equiv \int_{-\infty}^{\infty} dt \, \sgn(x-t) w^{(1)}(t) H_j \left( \frac{t}{\sqrt{2\tau}} \right).
\ee
Note that, for real $t$, we have $w^{(1)}(t) = \exp[-t^2/2(1+\tau)]$ from eq.\ (\ref{weightb1}).
In Appendix \ref{I_props} we derive some elementary properties of $I_j(x;\tau)$ which are used in the following.

On replacing the first occurence of $I_j(x;\tau)$ in eq.\ (\ref{GN_explicit}) using the recurrence relation eq.\ (\ref{Ij_recurrence_final}), we find that all except two of the terms involving $I_j(x;\tau)$ `telescope' down, to give
\begin{align}\label{GN_post_recurrence}
- \, \frac{G_N^{\RR\,(1)}(z_1,x_2)}{w^{(1)}(x_2)} &= \frac{1}{2\sqrt{2\pi}} \left\{ \sum_{j=0}^{N-2} \frac{1}{(j+1)!} \frac{\tau^{j+1}}{2^j} H_{j+1} \left( \frac{z_1}{\sqrt{2\tau}} \right) H_{j+1} \left( \frac{x_2}{\sqrt{2\tau}} \right) w^{(1)}(x_2) \right. \nn \\
 & \qquad \qquad \qquad + \frac{1}{(1+\tau)(N-1)!} \, \left( \frac{\tau}{2} \right)^{N-1/2} H_{N-1} \left( \frac{z_1}{\sqrt{2\tau}} \right) I_N(x_2;\tau) \nn \\
 & \qquad \qquad \qquad \qquad \left. {} - \frac{1}{1+\tau} \, \left( \frac{\tau}{2} \right)^{1/2} H_0 \left( \frac{z_1}{\sqrt{2\tau}} \right) I_1(x_2;\tau) \right\}.
\end{align}
Let us multiply both sides by $w^{(1)}(z_1)$, writing
\be
- \, \frac{w^{(1)}(z_1)}{w^{(1)}(x_2)} \, G_N^{\RR\,(1)}(z_1,x_2) = T_1(z_1,x_2) + T_2(z_1,x_2) + T_3(z_1,x_2)
\ee
where each $T_i$ corresponds to a separate line of eq.\ (\ref{GN_post_recurrence}).
The first term $T_1(z_1,x_2)$ is closely related to the $\beta=2$ kernel eq.\ (\ref{KNb2}) as follows:
\begin{align}\label{T1_lim}
T_1(z_1,x_2) &= \sqrt{\frac{\pi(1-\tau^2)}{2}} \,
\exp\left[ \frac{y_1^2}{1-\tau^2} \right] \, \left[ \erfc\left( \sqrt{\frac{2}{1-\tau^2}} \, |y_1| \right) \right]^{1/2}
\mcK_N^{(2)}(z_1,x_2) \nn \\
 & \qquad {} - \frac{1}{\sqrt{2\pi}} \, H_0 \left( \frac{z_1}{\sqrt{2\tau}} \right)H_0 \left( \frac{x_2}{\sqrt{2\tau}} \right) w^{(1)}(z_1) w^{(1)}(x_2) \nn \\
 &= \sqrt{\frac{\pi(1-\tau^2)}{2}} \,
\exp\left[ \frac{y_1^2}{1-\tau^2} \right] \, \left[ \erfc\left( \sqrt{\frac{2}{1-\tau^2}} \, |y_1| \right) \right]^{1/2}
\mcK_N^{(2)}(z_1,x_2) \nn \\
 & \qquad {} - \frac{1}{\sqrt{2\pi}} \,  w^{(1)}(z_1) w^{(1)}(x_2),
\end{align}
since $H_0(z)=1$. Using the fact that
\be
I_1(x;\tau) = - \frac{2\sqrt{2}(1+\tau)}{\sqrt{\tau}} w^{(1)}(x)
\ee
(which follows easily from the definition of $I_j(x;\tau)$ in eq.\ (\ref{Ij_def}) and the fact that $H_1(x) = 2x$), we can evaluate the third term $T_3(z_1,x_2)$ as
\be
T_3(z_1,x_2) = \frac{1}{\sqrt{2\pi}} w^{(1)}(z_1) w^{(1)}(x_2)
\ee
which cancels the second term in $T_1(z_1,x_2)$ in eq.\ (\ref{T1_lim}). Therefore, the final relation between the $\beta=1$ quantity $G_N^{\RR\,(1)}(z_1,x_2)$ and the $\beta=2$ kernel $\mcK_N^{(2)}(z_1,x_2)$ is:
\begin{align}\label{GN_from_KN_finite_N}
- \, \frac{w^{(1)}(z_1)}{w^{(1)}(x_2)} \, G_N^{\RR\,(1)}(z_1,x_2) &= \sqrt{\frac{\pi(1-\tau^2)}{2}} \, \exp\left[ \frac{y_1^2}{1-\tau^2} \right] \, \left[ \erfc\left( \sqrt{\frac{2}{1-\tau^2}} \, |y_1| \right) \right]^{1/2}
\mcK_N^{(2)}(z_1,x_2) \nn \\
 & \qquad {} + \frac{1}{2\sqrt{2\pi}(1+\tau)(N-1)!} \, \left( \frac{\tau}{2} \right)^{N-1/2} w^{(1)}(z_1) H_{N-1} \left( \frac{z_1}{\sqrt{2\tau}} \right) I_N(x_2;\tau).
\end{align}
We emphasise that the results in this section are exact, and are valid for even $N$ only.

\subsection{Definition of microscopic limit}

The limit of the $k$-point correlation function is defined as
\begin{align}
R_{k,\text{Ai}}^{(1)}(Z_1,\ldots,Z_k) & = \lim_{N\rightarrow\infty} \frac{1}{N^{k/3}} \, R_{k,N}^{(1)}(z_1,\ldots,z_k) \nn \\
 & = \lim_{N\rightarrow\infty} \frac{1}{N^{k/3}} \Pf_{i,j=1,\ldots,k} \left[ \left( \begin{array}{cc} \hat{\mcK}^{(1)}_N(z_i,z_j) & -G_N^{(1)}(z_i,z_j) \\ G_N^{(1)}(z_j,z_i) & -W_N^{(1)}(z_i,z_j) \end{array} \right) \right],
\end{align}
where the relationship between the $Z_i$ and the $z_i$ is the same as before, see eq.\ (\ref{edgelim}), and $\tau$ (implicit in the matrix elements) also scales with $N$ according to eq.\ (\ref{weaklim}).

In fact, the limits of the individual elements in this matrix do not exist. We can, however, use the definition of the Pfaffian to show that
\be\label{Pf_with_weight_adjusted_elements}
R_{k,\text{Ai}}^{(1)}(Z_1,\ldots,Z_k) = \Pf_{i,j=1,\ldots,k} \left[ \left( \begin{array}{cc} \hat{\mcK}_{\text{Ai}}^{(1)}(Z_i,Z_j) & -G_{\text{Ai}}^{(1)}(Z_i,Z_j) \\ G_{\text{Ai}}^{(1)}(Z_j,Z_i) & -W_{\text{Ai}}^{(1)}(Z_i,Z_j) \end{array} \right) \right],
\ee
with
\begin{align}
\hat{\mcK}_{\text{Ai}}^{(1)}(Z_1,Z_2) & \equiv \lim_{N\rightarrow\infty} \frac{1}{N^{1/3}} \, e^{-iN^{1/3}(Y_1+Y_2)} \, w^{(1)}(z_1)w^{(1)}(z_2) \hat{\mcK}_N^{(1)}(z_1,z_2), \label{K4_lim_def} \\
G_{\text{Ai}}^{(1)}(Z_1,Z_2) & \equiv \lim_{N\rightarrow\infty} \frac{1}{N^{1/3}} \, e^{-iN^{1/3}(Y_1-Y_2)} \, \frac{w^{(1)}(z_1)}{w^{(1)}(z_2)} \, G_N^{(1)}(z_1,z_2), \label{G4_lim_def} \\
W_{\text{Ai}}^{(1)}(Z_1,Z_2) & \equiv \lim_{N\rightarrow\infty} \frac{1}{N^{1/3}} \, e^{+iN^{1/3}(Y_1+Y_2)} \, \frac{1}{w^{(1)}(z_1)w^{(1)}(z_2)} \, W_N^{(1)}(z_1,z_2). \label{W4_lim_def}
\end{align}
These individual ``weight-adjusted'' limits do exist, and will now be derived explicitly.
In particular, note that the limit $\hat{\mcK}_{\text{Ai}}^{(1)}(Z_1,Z_2)$ of the pre-kernel
\textit{does} include the weight function.
Referring back to eqs.\ (\ref{G_split_R_C}) and (\ref{W_split_R_C}), we write the limiting matrix elements as separate terms, as follows:
\begin{align}
G_{\text{Ai}}^{(1)}(Z_1,Z_2) &= G_{\text{Ai}}^{\RR\,(1)}(Z_1,X_2) \, \delta(Y_2) + G_{\text{Ai}}^{\CC\,(1)}(Z_1,Z_2), \label{G4_lim_split} \\
W_{\text{Ai}}^{(1)}(Z_1,Z_2) &= W_{\text{Ai}}^{\RR\RR\,(1)}(X_1,X_2) \, \delta(Y_1)\delta(Y_2) + W_{\text{Ai}}^{\RR\CC\,(1)}(X_1,Z_2) \, \delta(Y_1) \nn \\
 & \qquad {} - W_{\text{Ai}}^{\RR\CC\,(1)}(X_2,Z_1) \, \delta(Y_2) + W_{\text{Ai}}^{\CC\CC\,(1)}(Z_1,Z_2) - \mcF_{\infty}^{(1)}(Z_1,Z_2). \label{W4_lim_split}
\end{align}
We devote a separate section to deriving each of these limits.

\subsubsection{Calculation of $\hat{\mcK}_{\text{Ai}}^{(1)}(Z_1,Z_2)$}

We begin by determining the limit in eq.\ (\ref{K4_lim_def}).
Consider first the weight function $w^{(1)}(z)$ given in eq.\ (\ref{weightb1_vs_b2}).
We take the large-$N$ limit of those factors in eq.\ (\ref{weightb1_vs_b2}) that depend on $y$ alone (as usual, under the scalings eqs.\ (\ref{weaklim}) and (\ref{edgelim})):
\be
\lim_{\substack{N\rightarrow\infty \\ \tau\rightarrow 1}} \exp\left[ \frac{y^2}{1-\tau^2} \right] \, \left[ \erfc\left( \sqrt{\frac{2}{1-\tau^2}} \,|y| \right) \right]^{1/2} = \exp\left[ \frac{Y^2}{2\sigma^2} \right] \, \sqrt{\erfc\left( \frac{|Y|}{\sigma} \right)},
\ee
leaving the $z$-dependent factors to be handled below.
On trivially rearranging the exact relationship eq.\ (\ref{prekernel_exact}), we can write the finite-$N$ $\beta=1$ pre-kernel $\hat{\mcK}_N^{(1)}(z_1,z_2)$ in terms of the $\beta=2$ pre-kernel $\hat{\mcK}_N^{(2)}(z_1,z_2)$, as follows:
\begin{align}\label{prekernel_relationship}
\hat{\mcK}_N^{(1)}(z_1,z_2) &= \frac{\sqrt{\pi}}{2\sqrt{2(1-\tau^2)}} \, (z_1 - z_2) \, \hat{\mcK}_{N-1}^{(2)}(z_1,z_2) \nn \\
 & \qquad {} -\frac{\tau^{N-1}}{2\sqrt{2\pi}(1-\tau^2)} \, (z_1-z_2) \sum_{j=0}^{N-2} \frac{1}{j!\,2^j} \, H_j \left( \frac{z_1}{\sqrt{2\tau}} \right)  H_j \left( \frac{z_2}{\sqrt{2\tau}} \right).
\end{align}
Multiplying eq.\ (\ref{prekernel_relationship}) throughout by the remaining $z$-dependent factors in the weight function, i.e.\ $\sqrt{w^{(2)}(z_1)w^{(2)}(z_2)}$, we see that the limit of the first term follows immediately from eq.\ (\ref{GinUE_edge_final}). The second term is in fact very similar to the first term, differing only in the fact that we have a factor $\tau^{N-1}$ outside the sum, instead of a factor $\tau^j$ inside. The effect of this is that we will get a result similar to eq.\ (\ref{GinUE_edge_final}), but without the $e^{\sigma^2 t}$ factor inside the integral.  Combining these, we therefore have
\begin{align}\label{K1_lim_final}
\hat{\mcK}_{\text{Ai}}^{(1)}(Z_1,Z_2) &= \frac{(Z_1-Z_2)}{4\sigma^2} \, \exp\left[ \frac{\sigma^6}{6} + \frac{\sigma^2(Z_1+Z_2)}{2} \right] \sqrt{\erfc\left(\frac{|Y_1|}{\sigma}\right)\erfc\left(\frac{|Y_2|}{\sigma}\right)} \nn \\
 & \qquad \times \int_0^{\infty} dt \, \big[ e^{\sigma^2 t} - 1 \big] \Ai\left(Z_1+\frac{\sigma^4}{4}+t\right) \Ai\left(Z_2+\frac{\sigma^4}{4}+t\right).
\end{align}

\subsubsection{Calculation of $G_{\text{Ai}}^{\RR\,(1)}(Z_1,X_2)$}

From eq.\ (\ref{G4_lim_def}), and given that
\be
\lim_{N\rightarrow\infty} \frac{\delta(y_2)}{N^{1/6}} = \delta(Y_2),
\ee
we now need to determine the limit
\be
G_{\text{Ai}}^{\RR\,(1)}(Z_1,X_2) \equiv \lim_{\substack{N\rightarrow\infty \\ \tau\rightarrow 1}} \frac{1}{N^{1/6}} \, e^{-iN^{1/3}Y_1} \, \frac{w^{(1)}(z_1)}{w^{(1)}(x_2)}  \, G_N^{\RR\,(1)}(z_1,x_2)
\ee
under our scalings.
Earlier, we saw in eq.\ (\ref{GN_from_KN_finite_N}) how to write the function $G_N^{\RR\,(1)}(z_1,x_2)$
in terms of $\mcK_N^{(2)}(z_1,x_2)$ and $I_N(x_2;\tau)$,
with $I_N(x;\tau)$ defined in eq.\ (\ref{Ij_def}) as an integral.
It is not immediately obvious whether the $N\rightarrow\infty$ limit operation and this integral commute,
and so we prefer to use the representation eq.\ (\ref{Ij_as_sum_of_Hermites}) for $I_N(x;\tau)$,
which does not involve any integral.
Eq.\ (\ref{Ij_as_sum_of_Hermites}) has two parts (one involving $I_0(x;\tau)$ and the other
involving a sum of Hermite polynomials),
and so (minus) $G_N^{\RR\,(1)}(z_1,x_2)$ can be written as a sum of three terms in total,
which we shall denote $U_1(Z_1,X_2)$, $U_2(Z_1,X_2)$ and $U_3(Z_1,X_2)$.
We take the limit of each term in turn.

For the first term from eq.\ (\ref{GN_from_KN_finite_N}), using eq.\ (\ref{weaklim}), we have
\begin{align}
U_1(Z_1,X_2) & \equiv \lim_{\substack{N\rightarrow\infty \\ \tau\rightarrow 1}} \frac{1}{N^{1/6}} \, e^{-iN^{1/3}Y_1} \, \sqrt{\frac{\pi(1-\tau^2)}{2}} \, \exp\left[ \frac{y_1^2}{1-\tau^2} \right] \, \left[ \erfc\left( \sqrt{\frac{2}{1-\tau^2}} \, |y_1| \right) \right]^{1/2}
 \mcK_N^{(2)}(z_1,x_2) \nn \\
 &= \sigma\sqrt{\pi} \lim_{\substack{N\rightarrow\infty \\ \tau\rightarrow 1}} \exp\left[ \frac{y_1^2}{1-\tau^2} \right] \, \left[ \erfc\left( \sqrt{\frac{2}{1-\tau^2}} \, |y_1| \right) \right]^{1/2}
 \frac{1}{N^{1/3}} \, e^{-iN^{1/3}Y_1} \, \mcK_N^{(2)}(z_1,x_2) \nn \\
 &= \sigma\sqrt{\pi} \, \exp\left[ \frac{Y_1^2}{2\sigma^2} \right] \, \sqrt{ \erfc\left( \frac{|Y_1|}{\sigma} \right) } \, \mcK_{\text{Ai}}^{(2)}(Z_1,X_2) \nn \\
 &= \exp\left[ \frac{\sigma^6}{6} + \frac{\sigma^2(Z_1+X_2)}{2} \right] \sqrt{ \erfc\left( \frac{|Y_1|}{\sigma} \right) } \int_0^{\infty} dt \, e^{\sigma^2 t} \, \Ai\left( Z_1 + \frac{\sigma^4}{4} + t \right)\, \Ai\left( X_2 + \frac{\sigma^4}{4} + t \right),
 \label{b1_U1_result}
\end{align}
where in the final step we used eq.\ (\ref{GinUE_edge_final}).

The second term comes from eqs.\ (\ref{GN_from_KN_finite_N}), (\ref{Ij_as_sum_of_Hermites}) and (\ref{I0_explicit}):
\begin{align}
U_2(Z_1,X_2) &\equiv \lim_{\substack{N\rightarrow\infty \\ \tau\rightarrow 1}} \frac{1}{N^{1/6}} \, e^{-iN^{1/3}Y_1} \, \frac{1}{2\sqrt{2\pi}(1+\tau)(N-1)!} \, \left( \frac{\tau}{2} \right)^{N-1/2} w^{(1)}(z_1) H_{N-1} \left( \frac{z_1}{\sqrt{2\tau}} \right) \nn \\
& \qquad \qquad \times \left( \frac{2}{\tau} \right)^{N/2} (N-1)!! \sqrt{2\pi(1+\tau)} \, \erf\left( \frac{x_2}{\sqrt{2(1+\tau)}} \right).
\end{align}
The error function tends to unity in this limit, and we handle the double factorial using eq.\ (\ref{double_fac_asymp_odd}) which gives
\be\label{double_factorial_asymptotic}
\lim_{\substack{N\rightarrow\infty \\ N \text{ even}}} \frac{(N-1)!!}{N^{1/4}\sqrt{(N-1)!}} = \left( \frac{2}{\pi} \right)^{1/4}.
\ee
Using eq.\ (\ref{weightb1_vs_b2}), we rewrite the weight function $w^{(1)}(z)$ in terms of the $\beta=2$ weight function $w^{(2)}(z)$.
Since $N^{1/12} \sim (N-1)^{1/12}$, this then gives
\begin{align}
U_2(Z_1,X_2) &= \lim_{\substack{N\rightarrow\infty \\ \tau\rightarrow 1}} \frac{1}{\sqrt{2(1+\tau)}} \exp\left[ \frac{y_1^2}{1-\tau^2} \right] \, \left[ \erfc\left( \sqrt{\frac{2}{1-\tau^2}} \, |y_1| \right) \right]^{1/2}  e^{-iN^{1/3}Y_1} h_{N-1}(z_1) \nn \\
 &= \frac{1}{2} \,  \exp\left[ \frac{\sigma^6}{12} + \frac{\sigma^2 Z_1}{2} \right] \, \sqrt{\erfc\left( \frac{|Y_1|}{\sigma} \right)} \Ai \left( Z_1 + \frac{\sigma^4}{4} \right),
 \label{b1_U2_result}
\end{align}
in which $h_j(z)$ was defined in eq.\ (\ref{beta_eq_2_hj_def}), and we used eq.\ (\ref{hjlim}) in the second step.
We note that $U_2(Z_1,X_2)$ actually depends on $Z_1$ alone, and so we will write it as $U_2(Z_1)$.

Finally, the third term is given by
\begin{align}
U_3(Z_1,X_2) &\equiv \lim_{\substack{N\rightarrow\infty \\ \tau\rightarrow 1}} \frac{1}{N^{1/6}} \, e^{-iN^{1/3}Y_1} \,\frac{1}{2\sqrt{2\pi}(1+\tau)(N-1)!} \, \left( \frac{\tau}{2} \right)^{N-1/2} w^{(1)}(z_1) H_{N-1} \left( \frac{z_1}{\sqrt{2\tau}} \right) \nn \\
 & \qquad \times \left( \frac{2}{\tau} \right)^{N/2} (N-1)!! (-2)(1+\tau)w^{(1)}(x_2) \sum_{k=0}^{N/2-1} \frac{1}{(2k+1)!!} \left( \frac{\tau}{2} \right)^{k+1/2} H_{2k+1}\left(\frac{x_2}{\sqrt{2\tau}}\right) \nn \\
 &= - \, \frac{U_2(Z_1)}{\sqrt{\pi}} \lim_{\substack{N\rightarrow\infty \\ \tau\rightarrow 1}} \sum_{\substack{k=1 \\ k\text{ odd}}}^{N-1} \frac{1}{k!!} \left( \frac{\tau}{2} \right)^{k/2} \sqrt{w^{(2)}(x_2)} \, H_k\left(\frac{x_2}{\sqrt{2\tau}}\right),
\end{align}
where we factored out the $Z_1$-dependency, and used that $w^{(1)}(x) = \sqrt{w^{(2)}(x)}$ for real $x$.
Now we follow a similar procedure to the inner sum of the $\beta=4$ case in eq.\ (\ref{b4_kernel_explicit}).
We change the summation variable from $k$ to $t$, and then allow $t$ to be continuous, redefining $k \equiv k(t,N)$ as
\be
k = \intodd(N + 1 - tN^{1/3}) \qquad \text{for} \qquad 0 \le t \le N^{2/3},
\ee
and writing the sum as an integral over $t$.
We again argue that we can take the limit inside the integral.
On handling the double factorial as before, eq.\ (\ref{double_fac_asymp_odd}),
we have
\begin{align}
U_3(Z_1,X_2) &= - U_2(Z_1) \int_0^{\infty} dt \, \lim_{\substack{N\rightarrow\infty \\ \tau\rightarrow 1}}
\Big( h_k(x_2) \Theta(k) \Big).
\end{align}
Therefore, using eq.\ (\ref{hjlim}), this immediately gives
\be
U_3(Z_1,X_2) = - U_2(Z_1) \, \exp\left[ \frac{\sigma^6}{12} + \frac{\sigma^2 X_2}{2} \right] \int_0^{\infty} dt \, e^{\sigma^2 t/2} \, \Ai\left( X_2 + \frac{\sigma^4}{4} + t \right).
\label{b1_U3_result}
\ee
We combine the three terms eqs.\ (\ref{b1_U1_result}), (\ref{b1_U2_result}) and (\ref{b1_U3_result}) to give
\begin{align}\label{G_as_U1_plus_U23}
- G_{\text{Ai}}^{\RR\,(1)}(Z_1,X_2) &= U_1(Z_1,X_2) + U_2(Z_1) + U_3(Z_1,X_2) \nn \\
  &= \exp\left[ \frac{\sigma^6}{6} + \frac{\sigma^2(Z_1+X_2)}{2} \right] \sqrt{\erfc\left( \frac{|Y_1|}{\sigma} \right)} \int_0^{\infty} dt \, e^{\sigma^2 t} \Ai\left( Z_1 + \frac{\sigma^4}{4} + t \right) \Ai\left( X_2 + \frac{\sigma^4}{4} + t \right) \nn \\
  & \qquad +  \tfrac{1}{2} \, \exp\left[ \frac{\sigma^6}{12} + \frac{\sigma^2 Z_1}{2} \right] \, \sqrt{\erfc\left( \frac{|Y_1|}{\sigma} \right)} \Ai \left( Z_1 + \frac{\sigma^4}{4} \right) \nn \\
  & \qquad \qquad \times \left\{ 1 -  \exp\left[ \frac{\sigma^6}{12} + \frac{\sigma^2 X_2}{2} \right] \int_0^{\infty} dt \, e^{\sigma^2 t/2} \Ai\left( X_2 + \frac{\sigma^4}{4} + t \right) \right\}.
\end{align}
The density of real eigenvalues is then given by
\be\label{Den_real_b1}
R_{1,\text{Ai}}^{\RR\,(1)}(Z) = -G_{\text{Ai}}^{\RR\,(1)}(X,X) \, \delta(Y)
\ee
where
\begin{align}
& -G_{\text{Ai}}^{\RR\,(1)}(X,X) = \exp\left[ \frac{\sigma^6}{6} + \sigma^2 X \right] \int_0^{\infty} dt \, e^{\sigma^2 t} \, \left[ \Ai\left( X + \frac{\sigma^4}{4} + t \right) \right]^2 \nn \\
  & \qquad +  \tfrac{1}{2} \, \exp\left[ \frac{\sigma^6}{12} + \frac{\sigma^2 X}{2} \right] \Ai \left( X + \frac{\sigma^4}{4} \right) \left\{ 1 -  \exp\left[ \frac{\sigma^6}{12} + \frac{\sigma^2 X}{2} \right] \int_0^{\infty} dt \, e^{\sigma^2 t/2} \Ai\left( X + \frac{\sigma^4}{4} + t \right) \right\}.
\end{align}

\subsubsection{Calculation of $G_{\text{Ai}}^{\CC\,(1)}(Z_1,Z_2)$, $W_{\text{Ai}}^{\RR\CC\,(1)}(X_1,Z_2)$ and $W_{\text{Ai}}^{\CC\CC\,(1)}(Z_1,Z_2)$}

These limits can all easily be expressed in terms of earlier results:
\begin{align}
G_{\text{Ai}}^{\CC\,(1)}(Z_1,Z_2) &= 2i \, \sgn(Y_2) \hat{\mcK}_{\text{Ai}}^{(1)}(Z_1,Z_2^*), \label{GC_final} \\
W_{\text{Ai}}^{\RR\CC\,(1)}(X_1,Z_2) &= 2i \, \sgn(Y_2) \, G_{\text{Ai}}^{\RR\,(1)}(Z_2^*,X_1), \\
W_{\text{Ai}}^{\CC\CC\,(1)}(Z_1,Z_2) &= -2i\,\sgn(Y_1) \, G_{\text{Ai}}^{\CC\,(1)}(Z_1^*,Z_2).
\end{align}
From eq.\ (\ref{GC_final}), the density of complex eigenvalues is given by
\begin{align}\label{Den_complex_b1}
R_{1,\text{Ai}}^{\CC\,(1)}(Z) &= -G_{\text{Ai}}^{\CC\,(1)}(Z,Z) \nn \\
 &= -2i \, \sgn(Y) \hat{\mcK}_{\text{Ai}}^{(1)}(Z,Z^*) \nn \\
 &= \frac{|Y|}{\sigma^2} \, \exp\left[ \frac{\sigma^6}{6} + \sigma^2 X \right] \erfc\left( \frac{|Y|}{\sigma} \right) \int_0^{\infty} dt \left[ e^{\sigma^2 t} - 1 \right] \left|\Ai\left( Z + \frac{\sigma^4}{4} + t \right)\right|^2,
\end{align}
where we used eq.\ (\ref{K1_lim_final}) in the final step.

\subsubsection{Calculation of $W_{\text{Ai}}^{\RR\RR\,(1)}(X_1,X_2)$}

From inserting eq.\ (\ref{GN_from_KN_finite_N}) into eq.\ (\ref{WNRR_expression}), we have at finite $N$
\begin{align}
& \frac{1}{w^{(1)}(x_1) w^{(1)}(x_2)} \, W_N^{\RR\RR\,(1)}(x_1,x_2) \nn \\
& = \intR dt \, \sgn(t-x_1) \left\{ \sqrt{\frac{\pi(1-\tau^2)}{2}}\, \mcK_N^{(2)}(t,x_2) + \frac{\left( \frac{\tau}{2} \right)^{N-1/2}}{2\sqrt{2\pi}(1+\tau)(N-1)!} \, H_{N-1}\left(\frac{t}{\sqrt{2\tau}}\right)I_N\left(x_2;\tau\right) \right\} \nn \\
& = - \, \frac{1}{\sqrt{2\pi}} \left\{ \sum_{j=0}^{N-1} \frac{1}{j!} \left( \frac{\tau}{2} \right)^j I_j(x_1;\tau) H_j\left( \frac{x_2}{\sqrt{2\tau}} \right) w^{(1)}(x_2) + \frac{ \left( \frac{\tau}{2} \right)^{N-1/2}}{2(1+\tau)(N-1)!} \, I_{N-1}(x_1;\tau)I_N(x_2;\tau)\right\}, \label{WNRR_as_sum}
\end{align}
using eq.\ (\ref{KNb2}) and the definition of $I_j(x;\tau)$ in eq.\ (\ref{Ij_def}).  Note that, despite appearances, this is anti-symmetric, i.e.\ $W_N^{\RR\RR\,(1)}(x_1,x_2) = - W_N^{\RR\RR\,(1)}(x_2,x_1)$.

As established in Appendix \ref{I_props},
each occurrence of $I_j(x;\tau)$ can be written as a sum over Hermite polynomials
(and an error function when $j$ is even),
and so this representation of $W_N^{\RR\RR\,(1)}(x_1,x_2)$ is essentially
a double sum of pairs of Hermite polynomials, plus some additional terms.
For convenience, we split the large-$N$ limit into several parts, writing:
\begin{align}
W_{\text{Ai}}^{\RR\RR\,(1)}(X_1,X_2) & \equiv \lim_{\substack{N\rightarrow\infty \\ \tau\rightarrow 1}}\frac{1}{w^{(1)}(x_1)w^{(1)}(x_2)} \, W_N^{\RR\RR\,(1)}(x_1,x_2) \nn \\
 &= - \, \frac{1}{\sqrt{2\pi}} \, \Big[ V_1(X_1,X_2) + V_2(X_1,X_2) + V_3(X_1,X_2) \Big],
\end{align}
where $V_1(X_1,X_2)$, $V_2(X_1,X_2)$ and $V_3(X_1,X_2)$ are the limits of the individual terms, defined as follows: $V_1$ contains the even $j$ terms in the sum in eq.\ (\ref{WNRR_as_sum}), $V_2$ the odd $j$ terms, and $V_3$ the final term that contains the product of $I_{N-1}$ and $I_N$.  Recall that $N$ is assumed even.

For $V_1(X_1,X_2)$ we have
\begin{align}
& V_1(X_1,X_2) \equiv \lim_{\substack{N\rightarrow\infty \\ \tau\rightarrow 1}} \sum_{\substack{j=0 \\ j\text{ even}}}^{N-2} \frac{1}{j!} \left( \frac{\tau}{2} \right)^j I_j(x_1;\tau) H_j\left( \frac{x_2}{\sqrt{2\tau}} \right) w^{(1)}(x_2) \nn \\
& = \lim_{\substack{N\rightarrow\infty \\ \tau\rightarrow 1}} \sum_{\substack{j=0 \\ j\text{ even}}}^{N-2} \frac{\left( \frac{\tau}{2} \right)^{j/2}}{j!!}  w^{(1)}(x_2) H_j\left( \frac{x_2}{\sqrt{2\tau}} \right) \left\{ \sqrt{2\pi(1+\tau)}- 2(1+\tau) \sum_{\substack{k=1\\k\text{ odd}}}^{j-1}\frac{\left(\frac{\tau}{2}\right)^{k/2}}{k!!} w^{(1)}(x_1) H_k\left(\frac{x_1}{\sqrt{2\tau}}\right) \right\} \nn \\
 & \equiv V_{11}(X_2) + V_{12}(X_1,X_2),
\end{align}
where in the second line we substituted using eq.\ (\ref{Ij_as_sum_of_Hermites})
and used the fact that the limit of the error function in eq.\ (\ref{I0_explicit}) is unity.
We now proceed in an almost identical way to the $\beta=4$ case, and so we can be brief.
\begin{align}
V_{11}(X_2) &\equiv \lim_{\substack{N\rightarrow\infty \\ \tau\rightarrow 1}}  \sqrt{2\pi(1+\tau)} \sum_{\substack{j=0 \\ j\text{ even}}}^{N-2} \frac{1}{j!!} \left( \frac{\tau}{2} \right)^{j/2} w^{(1)}(x_2) H_j\left( \frac{x_2}{\sqrt{2\tau}} \right) \nn \\
 & = \sqrt{2\pi} \, \exp\left[ \frac{\sigma^6}{12} + \frac{\sigma^2 X_2}{2} \right] \int_0^{\infty} dt \, e^{\sigma^2 t/2} \Ai\left( X_2 + \frac{\sigma^4}{4} + t \right).
\end{align}
For the double sum involved in $V_{12}$, we change the order of the summations as before, and arrive at
\begin{align}
V_{12}(X_1,X_2) & = - \, \sqrt{2\pi} \, \exp\left[ \frac{\sigma^6}{6} + \frac{\sigma^2 (X_1+X_2)}{2} \right]
\nn\\
&\quad\quad\quad\quad\times \int_0^{\infty} ds \, e^{\sigma^2 s/2} \Ai\left( X_1 + \frac{\sigma^4}{4} + s \right) \int_0^s dt \, e^{\sigma^2 t/2}\Ai\left( X_2 + \frac{\sigma^4}{4} + t \right).
\end{align}
By inspection, $V_2(X_1,X_2)$, which involves the odd $j$ terms in the sum, is seen to be equal to $V_{12}(X_1,X_2)$, and for $V_3(X_1,X_2)$ the result is
\begin{align}
V_3(X_1,X_2) & \equiv \lim_{\substack{N\rightarrow\infty \\ \tau\rightarrow 1}} \frac{1}{2(1+\tau)(N-1)!} \, \left( \frac{\tau}{2} \right)^{N-1/2} I_{N-1}(x_1;\tau)I_N(x_2;\tau) \nn \\
 & = - \, \sqrt{2\pi} \, \left\{ \exp\left[ \frac{\sigma^6}{12} + \frac{\sigma^2 X_1}{2} \right] \int_0^{\infty} dt \, e^{\sigma^2 t/2} \Ai\left( X_1 + \frac{\sigma^4}{4} + t \right) \right\} \nn \\
 & \qquad \qquad \qquad \times \left\{ 1 - \exp\left[ \frac{\sigma^6}{12} + \frac{\sigma^2 X_2}{2} \right] \int_0^{\infty} dt \, e^{\sigma^2 t/2} \Ai\left( X_2 + \frac{\sigma^4}{4} + t \right) \right\}.
\end{align}
Combining the components gives
\begin{align}\label{WRR_limit_final}
- W_{\text{Ai}}^{\RR\RR\,(1)}(X_1,X_2) &= - 2 A(X_1,X_2) + B(X_2) - B(X_1) \big(1 - B(X_2)\big) \nn \\
 & = A(X_2,X_1) - A(X_1,X_2) + B(X_2) - B(X_1),
\end{align}
where
\begin{align}
A(X_1,X_2) &= \exp\left[ \frac{\sigma^6}{6} + \frac{\sigma^2 (X_1+X_2)}{2} \right] \int_0^{\infty}\! ds \, e^{\sigma^2 s/2} \Ai\left( X_1 + \frac{\sigma^4}{4} + s \right)\! \int_0^s\! dt \, e^{\sigma^2 t/2} \Ai\left( X_2 + \frac{\sigma^4}{4} + t \right)\!\!, \nn \\
B(X) &= \exp\left[ \frac{\sigma^6}{12} + \frac{\sigma^2 X}{2} \right] \int_0^{\infty} dt \, e^{\sigma^2 t/2} \Ai\left( X + \frac{\sigma^4}{4} + t \right).
\end{align}

\subsubsection{Calculation of $\mcF_{\infty}^{(1)}(Z_1,Z_2)$}

Finally, we need to determine the limit of the bivariate $\beta=1$ weight function given in eq.\ (\ref{beta_eq_1_F_def}), which is straightforward.  We have:
\begin{align}
\mcF_{\infty}^{(1)}(Z_1,Z_2) & \equiv \lim_{N\rightarrow\infty} \frac{1}{N^{1/3}} \, \frac{1}{w^{(1)}(z_1)w^{(1)}(z_2)} \,\mcF^{(1)}(z_1,z_2) \nn \\
 & = 2i\delta^{(2)}(Z_1-Z_2^*)\sgn(Y_1) + \delta(Y_1)\delta(Y_2)\sgn(X_2-X_1),
\end{align}
where we used the following scalings of the Dirac delta function:
\be
\frac{\delta(y)}{N^{1/6}} = \delta(Y) \qquad \text{and} \qquad \frac{\delta^{(2)}(z_1-z_2^*)}{N^{1/3}} = \delta^{(2)}(Z_1-Z_2^*).
\ee

\subsection{Hermitian and strongly non-Hermitian limits}
\label{beta_eq_1_density_Hermitian}

We consider the Hermitian limit of the microscopic eigenvalue density in this section, with the limits of the more general $k$-point correlation functions being relegated to Appendix \ref{Hermlimb1}.
For the density of real eigenvalues, the Hermitian limit $\sigma\rightarrow 0$ is quite straightforward. We can simply set $\sigma = 0$ in eq.\ (\ref{Den_real_b1}), to give
\be
\lim_{\sigma\rightarrow 0} R_{1,\text{Ai}}^{\RR\,(1)}(Z) = \left\{ \int_0^{\infty} dt \, \left[ \Ai( X + t) \right]^2 + \tfrac12 \Ai(X) \left( 1 - \int_0^{\infty} dt \, \Ai(X + t) \right) \right\} \, \delta(Y).
\ee
For the density of complex eigenvalues, we take eq.\ (\ref{Den_complex_b1}), and expand the factor $e^{\sigma^2 t}$ inside the integral in powers of $\sigma^2 t$, to give at small $\sigma$
\be
R_{1,\text{Ai}}^{\CC\,(1)}(Z) = |Y| \exp\left[ \frac{\sigma^6}{6}  + \sigma^2 X \right] \erfc\left( \frac{|Y|}{\sigma} \right) \int_0^{\infty}\! dt\left[ \frac{\left( 1 + \sigma^2 t + \tfrac12 \sigma^4 t^2 + \ldots \right)-1}{\sigma^2} \right] \left| \Ai\left(Z + \frac{\sigma^4}{4} + t\right) \right|^2\!.
\ee
Because
\be
\lim_{\sigma\rightarrow 0} |Y| \, \erfc\left( \frac{|Y|}{\sigma} \right) = 0 \qquad \forall \,\, Y \in \RR
\ee
and the other factors have a finite limit,
the $\sigma\rightarrow 0$ limit of $R_{1,\text{Ai}}^{\CC\,(1)}(Z)$ is clearly zero for all $Z$.
This is indeed the anticipated behaviour;
the (average) number of complex eigenvalues gradually decreases as $\sigma$ gets closer to zero.
For small, but finite, $\sigma$, most eigenvalues (on average) already lie exactly on the real axis.
So qualitatively, the $\beta=1$ case is entirely different from what happens with $\beta=2$ or $\beta=4$.

We now turn to the strongly non-Hermitian limit $\sigma\to\infty$.
For the various elements, we have
\begin{align}
\hat{\mcK}_{\text{edge}}^{(1)}(\hat{Z}_1,\hat{Z}_2) &\equiv \lim_{\sigma\rightarrow\infty} 2 \sigma^2 \hat{\mcK}_{\text{Ai}}^{(1)}(\sigma\hat{Z}_1,\sigma\hat{Z}_2) \nn \\
 &= \frac{\hat{Z}_1-\hat{Z}_2}{8\sqrt{\pi}} \sqrt{\erfc(|\hat{Y}_1|)\erfc(|\hat{Y}_2|)}  \,  \exp\left[ - \, \frac{(\hat{Z}_1-\hat{Z}_2)^2}{4} \right] \erfc\left( \frac{\hat{Z}_1+\hat{Z}_2}{2} \right), \\
G_{\text{edge}}^{\CC\,(1)}(\hat{Z}_1,\hat{Z}_2) &\equiv \lim_{\sigma\rightarrow\infty} 2 \sigma^2 G_{\text{Ai}}^{\CC\,(1)}(\sigma\hat{Z}_1,\sigma\hat{Z}_2) = 2i \, \sgn(\hat{Y}_2) \hat{\mcK}_{\text{edge}}^{(1)}(\hat{Z}_1,\hat{Z}_2^*), \\
- G_{\text{edge}}^{\RR\,(1)}(\hat{Z}_1,\hat{X}_2) &\equiv - \lim_{\sigma\rightarrow\infty} 2 \sigma G_{\text{Ai}}^{\RR\,(1)}(\sigma\hat{Z}_1,\sigma\hat{X}_2) \nn \\
 &= \frac{1}{2\sqrt{\pi}} \sqrt{\erfc(|\hat{Y}_1|)} \, \exp\left[ {} - \frac{(\hat{Z}_1-\hat{X}_2)^2}{4} \right] \erfc\left( \frac{\hat{Z}_1+\hat{X}_2}{2} \right) \nn \\
 & \qquad + \frac{1}{\sqrt{2\pi}} \sqrt{\erfc(|\hat{Y}_1|)} \, \exp\left[ {} - \frac{\hat{Z}_1^2}{2} \right] \left\{ 1 - \frac{1}{2}\erfc\left( \frac{\hat{X}_2}{\sqrt{2}} \right) \right\}, \\
W_{\text{edge}}^{\CC\CC\,(1)}(\hat{Z}_1,\hat{Z}_2) &\equiv \lim_{\sigma\rightarrow\infty} 2 \sigma^2 W_{\text{Ai}}^{\CC\CC\,(1)}(\sigma\hat{Z}_1,\sigma\hat{Z}_2) = 4 \, \sgn(\hat{Y}_1) \sgn(\hat{Y}_2) \hat{\mcK}_{\text{edge}}^{(1)}(\hat{Z}_1^*,\hat{Z}_2^*), \\
W_{\text{edge}}^{\RR\CC\,(1)}(\hat{X}_1,\hat{Z}_2) &\equiv \lim_{\sigma\rightarrow\infty} 2 \sigma W_{\text{Ai}}^{\RR\CC\,(1)}(\sigma\hat{X}_1,\sigma\hat{Z}_2) =
-2i \, \sgn(\hat{Y}_2) G_{\text{edge}}^{\RR\,(1)}(\hat{Z}_2^*,\hat{X}_1), \\
- W_{\text{edge}}^{\RR\RR\,(1)}(\hat{X}_1,\hat{X}_2) &\equiv \lim_{\sigma\rightarrow\infty} - 2 W_{\text{Ai}}^{\RR\RR\,(1)}(\sigma\hat{X}_1,\sigma\hat{X}_2) \nn \\
 &= P(\hat{X}_1,\hat{X}_2) - P(\hat{X}_2,\hat{X}_1) + Q(\hat{X}_2) - Q(\hat{X}_1),
\end{align}
where
\begin{align}
P(\hat{X}_1,\hat{X}_2) &= \frac{1}{\sqrt{2\pi}} \int_0^{\infty} ds \, \exp\left[ - \, \frac{(\hat{X}_2+s)^2}{2} \right]  \erf\left( \frac{\hat{X}_1+s}{\sqrt{2}} \right), \\
Q(\hat{X}) &= \tfrac12 \, \erfc\left( \frac{\hat{X}}{\sqrt{2}} \right).
\end{align}
The densities of complex and real eigenvalues are therefore given by
\begin{align}
R^{{\mathbb C}(1)}_{1,\,\text{edge}}(\hat{Z}) &= -2i \, \sgn(\hat{Y}) \hat{\mcK}_{\text{edge}}^{(1)}(\hat{Z},\hat{Z}^*)
= \frac{|\hat{Y}|}{2\sqrt{\pi}}
\ \text{erfc}(|\hat{Y}|)\ \exp[\hat{Y}^2]\ \erfc(\hat{X})\ ,  \\
R^{{\mathbb R}(1)}_{1,\,\text{edge}}(\hat{X}) &= \frac{1}{2\sqrt{\pi}}\left\{
\erfc(\hat{X})+\frac{1}{\sqrt{2}}
\exp\left[- \, \frac{\hat{X}^2}{2}\right]\ \erfc\left(- \, \frac{\hat{X}}{\sqrt{2}}\right)\right\},
\end{align}
the latter agreeing with eq.\ 5.13 of the second paper of ref.\ \cite{Forrester08}, after a trivial rescaling of the argument $\hat{X}$.

Let us now consider the case of large $\hat{Y}$, i.e.\ far from the real axis which is ``special'' for this
ensemble.
Since for large, real $y$
\be
\erfc(|y|) \sim \frac{1}{\sqrt{\pi}|y|} \, e^{-y^2},
\ee
we have at large $|\hat{Y}|$
\be
\hat{\mcK}_{\text{edge}}^{(1)}(\hat{Z}_1,\hat{Z}_2) \sim \frac{\hat{Z}_1 - \hat{Z}_2}{8\pi\sqrt{|\hat{Y}_1\hat{Y}_2|}} \, \exp\left[ - \, \frac{\hat{Y}_1^2+\hat{Y}_2^2}{2} - \frac{(\hat{Z}_1-\hat{Z}_2)^2}{4} \right] \erfc\left( \frac{\hat{Z}_1+\hat{Z}_2}{2} \right),
\ee
and so for the density we have
\be
R^{{\mathbb C}(1)}_{1,\,\text{edge}}(\hat{Z}) \sim \frac{1}{2{\pi}} \erfc(\hat{X}),
\ee
recovering the same result as for $\beta=2$ and $\beta=4$, see \cite{BorSin}, thereby demonstrating the universality
of this quantity.

As a final check, we turn again to the limit in the vicinity of the eigenvalue
with the largest real part. Due to the known dominance of the complex eigenvalues over the real eigenvalues we will focus here on the limit of the kernel $\hat{\mcK}_{\text{Ai}}^{(1)}(Z_1,Z_2)$, eq.\ (\ref{K1_lim_final}).
Apart from the $Y$-dependence this kernel is very similar to the $\beta=2$ kernel in eq.\ (\ref{GinUE_edge_final}), the limit of which has been discussed in detail already.
Therefore we can be brief. We define in analogy to eq.\ (\ref{Mbeta2})
\begin{align}
M^{(1)}(z_1,z_2) &\equiv \lim_{\sigma\rightarrow\infty}
\sigma a(\sigma)b(\sigma) \,
\exp\left[ \frac{iX_1Y_1}{\sigma} + \frac{iY_1^3}{3\sigma^3} + \frac{iX_2Y_2}{\sigma} + \frac{iY_2^3}{3\sigma^3} \right] \,  \hat{\mcK}_{\text{Ai}}^{(1)}(Z_1,Z_2;\sigma)\nn\\
&=
\frac{1}{\sqrt{\pi}} \, \exp\left[ - \frac{x_1+x_2+y_1^2+y_2^2}{2} \right]
\lim_{\sigma\rightarrow\infty}\left[
\frac{\sqrt{\pi}}{4\sigma}[a(\sigma)(x_1-x_2)+i\sigma b(\sigma)(y_1-y_2)]\right.
\nn\\
 &  \qquad \times\left.
\sqrt{\erfc(b(\sigma)|y_1|)\erfc(b(\sigma)|y_2|)}
\,e^{b(\sigma)^2(y_1^2+y_2^2)}
\left[1 + \frac{i(y_1+y_2)\sigma^{3/2}}{(6\log\sigma)^{3/4}}\right]^{-1}\right]
\nn\\
&= \frac{i(y_1-y_2)}{4\sqrt{\pi|y_1y_2|}}
\, \exp\left[ - \frac{x_1+x_2+y_1^2+y_2^2}{2} \right]\delta_{y_1,-y_2} = \frac{i}{2} \, \sgn(y_1) M^{(2)}(z_1,z_2) \ ,
\end{align}
with the scalings given in eqs.\ (\ref{XYscale}) and (\ref{abc_def}). The result has the same form as the $\beta=2$ Poisson kernel.
We note that the density of complex eigenvalues is obtained from eq.\ (\ref{Den_complex_b1})
by multiplication of the limiting kernel with complex conjugate arguments by the factor $-2i\,\sgn(y)$,
thereby giving the same density.
The Poisson kernel above relates in precisely the same way to the Gumbel distribution
as discussed for $\beta=2$ and $4$.
This would appear to be a new result,
although again it parallels the corresponding (recent) result for the eigenvalue
with largest modulus, see the final reference of \cite{Rider}.

\sect{Conclusions}\label{conclusions}

In this paper we have derived the interpolating Airy matrix-kernels
for the $\beta=1$ and $\beta=4$ symmetry classes of non-Hermitian random matrices.
This was achieved by taking the weakly non-Hermitian large-$N$ limit
of the corresponding elliptical Ginibre ensembles (which are Gaussian)
whilst simultaneously magnifying the vicinity of the largest (or smallest) real eigenvalue.

As a preliminary, we rederived the interpolating Airy kernel for the $\beta=2$ case
by using a rather straightforward asymptotic expansion of the Hermite polynomials in the complex plane.
This is distinct from the saddle-point analysis of a complex integral representation
of the kernel of these polynomials that was used in the original derivation in \cite{Bender}.
Our subsequent analysis for  $\beta=1$ and $\beta=4$ is non-trivial, especially for the $\beta=1$ case
where it is known that the large-$N$ limit does not always commute
with the integral expressions for the matrix kernel at finite $N$,
as in the case of the chiral GOE, for example.

We performed consistency checks of our results for the matrix-kernel elements
for $\beta=1$ and $\beta=4$ by taking both the Hermitian limit and two distinct strongly non-Hermitian limits.
In particular, the Hermitian limit leads to the universal result for the well known corresponding Airy kernels. At strong non-Hermiticity one can either magnify the region around the edge of the spectrum, leading to the complementary error function kernel,
or one can study the vicinity of the eigenvalue with the largest real part, leading to the Poisson kernel and thus ultimately to the Gumbel distribution.

We found agreement in all cases where the limiting correlation functions were known.
For $\beta=4$ in the strongly non-Hermitian limit at the spectral edge,
we were unable to find the result in the existing literature.
However, our limiting results are consistent with the hypothesis that the
eigenvalue density is universal for all three values of $\beta$.

Two open problems in particular seem to us to be worth pursuing.
First, to find rigorous proofs (or even heuristic arguments) in favour of the universality
of these one-parameter deformations of Hermitian random matrix ensembles.
And second, to find interesting physics applications for our kernels,
where possible candidates might include growth processes in higher dimensions.

\noindent
{\bf Acknowledgements:}
Partial support by the SFB$|$TR12 ``Symmetries and Universality
in Mesoscopic Systems'' of the German research council DFG is acknowledged (G.A.).
We would like to thank Queen Mary University of London (G.A.) and Bielefeld University (M.J.P.) for hospitality.

\appendix
\sect{The deformed Airy function}
\label{DeformedAiryFunction}

In this appendix we introduce the deformed Airy function
\be\label{Aid_def}
\Aid(Z,\sigma) \equiv \exp\left[ \frac{\sigma^6}{12} + \frac{\sigma^2 Z}{2} \right] \Ai\left( Z + \frac{\sigma^4}{4} \right),
\ee
and consider three of its scaling limits.
Using the large argument asymptotic behaviour of the Airy function (see eq.\ 10.4.59 of \cite{Abram})
\begin{align}
\Ai(w) &\sim \frac{\exp\left[ {} - \tfrac23 w^{3/2} \right]}{2\sqrt{\pi}w^{1/4}} \left( 1 - \frac{5}{48w^{3/2}} + \mcO(w^{-3}) \right) \nn \\
 &= \frac{1}{2\sqrt{\pi}} \, \exp\left[ {} - \tfrac23 w^{3/2} - \tfrac14 \log w  + \log\left( 1 - \frac{5}{48w^{3/2}} + \mcO(w^{-3}) \right) \right]
\end{align}
we can easily show that, at large $\sigma$ and fixed $Z$ (or, at least, for $Z$ growing more slowly than $\sigma^4$),
we have
\be\label{Aid_limit}
\Aid(Z,\sigma) = \frac{1}{\sqrt{2\pi}\,\sigma} \exp\left[ {} - \, \frac{Z^2}{2\sigma^2} - \frac{Z}{\sigma^4} + \frac{2Z^3-5}{6\sigma^6} + \ldots \right].
\ee
From this, we immediately have the first of our scaling limits,
\be
\lim_{\sigma\rightarrow\infty} \sigma \, \Aid(\sigma u, \sigma) = \frac{1}{\sqrt{2\pi}} \, e^{-u^2/2}
\ee
where we scale the first argument proportional to $\sigma$.

We now consider a second scaling limit.
In our interpolating Airy kernels, e.g.\ eqs.\ (\ref{GinUE_edge_final}) and (\ref{beta_eq_4_K_final}),
the deformed Airy function typically appears in combination with an additional factor:
\be
f(Z,t,\sigma) \equiv \exp\left[ - \frac{Y^2}{2\sigma^2} \right] \Aid(Z+t, \sigma).
\ee
Consider therefore the behaviour of
\begin{align}
g(Z,t,\sigma) &\equiv f(X + i\sigma Y, t, \sigma) \nn \\
 &= \exp\left[ - \, \frac{Y^2}{2} \right] \, \Aid(X + i\sigma Y + t, \sigma),
\end{align}
where it should be noted that we are scaling the real and imaginary parts of the argument $Z$ differently.
At large $\sigma$ and fixed $X$ and $Y$ (or, at least, with $X$ and $Y$ growing sufficiently slowly with $\sigma$), we then have from eq.\ (\ref{Aid_limit})
\begin{align}\label{g_def}
g(Z,t,\sigma) &= \frac{1}{\sqrt{2\pi}\,\sigma} \,  \exp\left[ - \frac{Y^2}{2} \right] \, \exp\left[ - \frac{(X + i\sigma Y + t)^2}{2\sigma^2} - \frac{X + i\sigma Y + t}{\sigma^4} + \frac{2(X + i\sigma Y + t)^3-5}{6\sigma^6} + \ldots \right] \nn \\
 &= \exp\left[ - \frac{iXY}{\sigma} - \frac{iY^3}{3\sigma^3} \right] \, h(Z,t,\sigma)
\end{align}
where
\begin{align}\label{h_def}
h(Z,t,\sigma) &= \frac{1}{\sqrt{2\pi}\,\sigma} \,  \exp\left[ \left( {} - \frac{X^2}{2\sigma^2} - \frac{iY}{\sigma^3} - \frac{X}{\sigma^4} - \frac{Y^2X}{\sigma^4} + \frac{iX^2Y}{\sigma^5} + \frac{X^3}{3\sigma^6} - \frac{5}{6\sigma^6} + \ldots \right) \right. \nn \\
 & \qquad \qquad \qquad {} - t \left( \frac{iY}{\sigma} + \frac{X}{\sigma^2} + \frac{1}{\sigma^4} + \frac{Y^2}{\sigma^4} - \frac{2iXY}{\sigma^5} - \frac{X^2}{\sigma^6} + \ldots \right) \nn \\
 & \qquad \qquad \qquad \qquad \left.  {} - t^2 \left( \frac{1}{2\sigma^2} - \frac{iY}{\sigma^5} - \frac{X}{\sigma^6} + \ldots \right) + \frac{t^3}{3\sigma^6} + \ldots \right].
\end{align}
In the final expression here,
we have only explicitly written those terms that come from expanding
the terms in the exponent in eq.\ (\ref{g_def}) up to cubic order.
We have ordered the terms in eq.\ (\ref{h_def}) by increasing powers of $t$,
and then by decreasing powers of $\sigma$,
although shortly we will let $X$, $Y$ and $t$ themselves all be dependent on $\sigma$,
and so the relative sizes of the terms will change.
Note that we chose to factor out two terms (the difference between $g$ and $h$) for later convenience.

Let us now introduce some new coordinates $x$ and $y$, where again we scale the real and imaginary parts in different ways:
\begin{align}
\label{XYscale}
X &= a(\sigma)x + c(\sigma), \nn \\
Y &= b(\sigma)y,
\end{align}
in which the scalings and shift are given by (see \cite{Bender})
\begin{align}\label{abc_def}
a(\sigma) &\equiv \frac{\sigma}{\sqrt{6\log\sigma}}, \nn \\
b(\sigma) &\equiv \frac{\sigma^{3/2}}{(6\log\sigma)^{1/4}}, \nn \\
c(\sigma) &\equiv a(\sigma)\Big( 3 \log\sigma - \tfrac54 \log(6\log\sigma) - \log(2\pi) \Big).
\end{align}
We can choose how to scale $t$, since this will be an integration variable
in the expressions for the limiting kernels.
In order to make the analysis as straightforward as possible,
it turns out that the following, simple scaling is optimal:
\be
t = d(\sigma) u,
\ee
where
\be\label{d_def}
d(\sigma) \equiv \sigma.
\ee
So now consider the function
\begin{align}\label{m_def}
m(x,y,u,\sigma) &\equiv h(X + iY, t, \sigma) \nn \\
 &= h\Big(a(\sigma)x + c(\sigma) + ib(\sigma)y, d(\sigma)u, \sigma\Big).
\end{align}
On substituting into eq.\ (\ref{h_def}), we get a \textit{very} large number of terms in the exponent
(almost 80, in fact).
However, most of these will vanish in the limit $\sigma\rightarrow\infty$,
so we will retain only those that do not.
After some simplification, we arrive at our second scaling limit of the deformed Airy function,
\begin{align}\label{m_formula}
m(x,y,u,\sigma) &\sim \frac{(6 \log\sigma)^{5/8}}{\sigma^{7/4}}
\exp\left[ {} - \frac{x+y^2}{2} - \frac{u^2}{2} - \frac{u\sqrt{6\log\sigma}}{2} - \frac{iuy\sigma^{3/2}}{(6\log\sigma)^{1/4}} \right].
\end{align}

For the third scaling limit, we use the asymptotic behaviour of the Airy function eq.\ 10.4.60 of \cite{Abram}
\be
\Ai(-w) \sim \frac{\sin\left( \tfrac23 w^{3/2} + \frac{\pi}{4} \right)}{\sqrt{\pi} \, w^{1/4}}
\ee
for large positive $w$ to show that, for $\alpha>0$ and at large real $w$, (with $z$, $\sigma$ and $\alpha$ fixed)
\begin{align}\label{Aid_bulk_limit}
\Aid\left( \frac{z}{w} - \alpha w^2, \frac{\sigma}{w} \right) &\sim
\frac{\exp[-\alpha\sigma^2/2]}{\sqrt{\pi} \, \alpha^{1/4}\sqrt{w}} \, \sin \left( \tfrac23 \alpha^{3/2} w^3 \left( 1 - \frac{z}{\alpha w^3} - \frac{\sigma^4}{4\alpha w^6} \right)^{3/2} + \frac{\pi}{4} \right) \nn \\
 &\sim \frac{\exp[-\alpha\sigma^2/2]}{\sqrt{\pi} \, \alpha^{1/4}\sqrt{w}} \, \sin \left( \tfrac23 \alpha^{3/2} w^3 - \sqrt{\alpha} \, z + \frac{\pi}{4} \right).
\end{align}

\sect{Hermitian limit of the $\beta=4$ matrix-kernel}
\label{Hermlimb4}

As a check in this appendix we will take the Hermitian limit ${\sigma\to0}$ of the $\beta=4$ matrix-kernel elements eq.\ (\ref{beta_eq_4_K_final}) as they appear in eq.\ (\ref{Rkb4alt}).
Since all the elements of the matrix-kernel coincide to leading order,
this leads to the appearance of first and second order derivatives acting on the Hermitian limit of the kernel.
The same problem was encountered previously when taking the Hermitian limit of the $\beta=4$ kernel
\cite{ABa} in the microscopic hard-edge scaling limit, starting from the (interpolating) Bessel kernel in the complex plane \cite{A05}. Because we will follow \cite{ABa} closely, we can be brief.
There it was shown by using properties of the Pfaffian that the matrix-kernel inside eq.\ (\ref{Rkb4alt})
can be changed as follows, without changing the value of the Pfaffian:
\begin{align}
\label{Pf-shift}
&\left(
\begin{array}{ll}
\mcK(z_i,z_j) & \mcK(z_i,z_j^\ast) \\
\mcK(z_i^\ast,z_j) & \mcK(z_i^\ast,z_j^\ast)
\end{array}
\right)
\rightarrow\\
&\left(
\begin{array}{ll}
  \mcK(z_i,z_j) - \mcK(z_i,z_j^\ast) -
  \mcK(z_i^\ast ,z_j) + \mcK(z_i^\ast, z_j^\ast)) &
\frac{1}{2}(  \mcK(z_i,z_j)- \mcK(z_i^\ast ,z_j) +
\mcK(z_i,z_j^\ast) - \mcK(z_i^\ast, z_j^\ast) )
\\
\frac{1}{2}(  \mcK(z_i,z_j) - \mcK(z_i,z_j^\ast) +
\mcK(z_i^\ast ,z_j) - \mcK(z_i^\ast, z_j^\ast))  &
\frac{1}{4}(  \mcK(z_i,z_j) + \mcK(z_i,z_j^\ast) +
  \mcK(z_i^\ast ,z_j) + \mcK(z_i^\ast, z_j^\ast))
\end{array}
\right)\!.
\nn
\end{align}
We have suppressed all labels here, partly for simplicity, and partly because the replacement is an identity both before and after taking the large-$N$ limit. We will now expand the matrix-kernel elements from eq.\ (\ref{beta_eq_4_K_final}) as in subsection \ref{beta_eq_4_density_Hermitian}, but this time to linear order in both $Y_1$ and $Y_2$, with $Y_{1,2},\sigma\ll1$.
\begin{align}
\label{beta_eq_4_G_Hermitian_approx}
\mcK_{\text{Ai}}^{(4)}(Z_1,Z_2^*)
 & \approx  \frac{-i\sqrt{|Y_1Y_2|}}{4\sqrt{\pi}\,\sigma^3} \, \exp\left[ - \, \frac{Y_1^2+Y_2^2}{2\sigma^2} \right] \nn \\
 & \qquad \times \left\{ \int_0^{\infty} ds \int_0^s dt \, \Big( \Ai(X_2+s)-iY_2\Ai\hspace{0mm}'(X_2+s) \Big) \Big( \Ai(X_1+t)+iY_1\Ai\hspace{0mm}'(X_1+t) \Big) \right. \nn \\
 & \qquad \qquad \qquad \qquad \left. {} - \Big( \Ai(X_1+s)+iY_1\Ai\hspace{0mm}'(X_1+s) \Big) \Big( \Ai(X_2+t)-iY_2\Ai\hspace{0mm}'(X_2+t) \Big) \right\} \nn \\
 & \equiv  v(Y_1,Y_2)\, \Big[ T_1(X_1,X_2) + iY_1T_2(X_1,X_2) + iY_2T_3(X_1,X_2) + Y_1Y_2T_4(X_1,X_2) \Big],
\end{align}
where
\begin{align}
v(Y_1,Y_2)&=\frac{-i\sqrt{|Y_1Y_2|}}{4\sqrt{\pi}\,\sigma^3} \, \exp\left[ - \, \frac{Y_1^2+Y_2^2}{2\sigma^2} \right],\nn\\
T_1(X_1,X_2) & = \int_0^{\infty} ds \int_0^s dt \, \Big[ \Ai(X_2+s)\Ai(X_1+t) - \Ai(X_1+s)\Ai(X_2+t) \Big], \nn \\
T_2(X_1,X_2) & = \int_0^{\infty} ds \int_0^s dt \, \Big[ \Ai(X_2+s)\Ai\hspace{0mm}'(X_1+t) - \Ai\hspace{0mm}'(X_1+s)\Ai(X_2+t) \Big] \nn \\
 & = 2\int_0^{\infty} ds \, \Ai(X_1+s)\Ai(X_2+s) -\Ai(X_1) \int_0^{\infty} ds \, \Ai(X_2+s)\ , \nn \\
T_3(X_1,X_2) & = \int_0^{\infty} ds \int_0^s dt \, \Big[ -\Ai\hspace{0mm}'(X_2+s)\Ai(X_1+t) + \Ai(X_1+s)\Ai\hspace{0mm}'(X_2+t) \Big] \nn \\
 & =  2\int_0^{\infty} ds \, \Ai(X_1+s)\Ai(X_2+s) -\Ai(X_2) \int_0^{\infty} ds \, \Ai(X_1+s)\ ,\nn \\
T_4(X_1,X_2) & =\int_0^{\infty} ds \int_0^s dt \, \Big[ \Ai\hspace{0mm}'(X_2+s)\Ai\hspace{0mm}'(X_1+t) - \Ai\hspace{0mm}'(X_1+s)\Ai\hspace{0mm}'(X_2+t) \Big] \nn \\
 & = \int_0^{\infty} ds \, \Big[ \Ai(X_1+s)\Ai\hspace{0mm}'(X_2+s) - \Ai(X_2+s)\Ai\hspace{0mm}'(X_1+s) \Big].
\label{Tdefs}
\end{align}
Whilst the integration in $T_4$ is straightforward,
in $T_2$ and $T_3$ we have again used eq.\ (\ref{AiryInt}).
Applying the shift eq.\ (\ref{Pf-shift}) under the Pfaffian we obtain
\begin{align}
\lim_{\sigma,Y\ll 1}R_{k,\text{Ai}}^{(4)}(Z_1,\ldots,Z_k) &\approx
\Pf_{i,j=1,\ldots,k}
\left[\left(
\begin{array}{cc}
-4Y_iY_jv(Y_i,Y_j)T_4(X_i,X_j)
&2iY_iv(Y_i,Y_j)T_2(X_i,X_j)\\
-2iY_jv(Y_i,Y_j)T_3(X_i,X_j)
&v(Y_i,Y_j)T_1(X_i,X_j)\\
\end{array}
\right)\right]\nn\\
&=\prod_{j=1}^k\frac{2Y_j^2}{\sqrt{\pi}\,\sigma^3} \, \exp\left[ - \, \frac{Y_j^2}{\sigma^2} \right]
\Pf_{i,j=1,\ldots,k}
\left[\left(
\begin{array}{cc}
-\tfrac14 T_4(X_i,X_j)
&\tfrac14 T_2(X_i,X_j)\\
-\tfrac14 T_3(X_i,X_j)
&-\tfrac14 T_1(X_i,X_j)\\
\end{array}
\right)\right]
\ .\nn\\
\end{align}
In the second step we have taken out all the $Y$-dependent factors, which, using eq.\ (\ref{deltadef2}),
will give rise to the product $\displaystyle \prod_{j=1}^k\delta(Y_j)$ in the limit $\sigma\to0$ as expected.
To see the matching with the known result for $\beta=4$ with real eigenvalues we note that
\be
T_2(X_1,X_2)=T_3(X_2,X_1)=2\int_0^{\infty} ds \, \Ai(X_1+s)\Ai(X_2+s) -\Ai(X_1) \int_{X_2}^{\infty} ds \, \Ai(s)
\ee
where the first part is proportional to the $\beta=2$ Airy kernel of real eigenvalues, eq.\ (\ref{KAib2lim}). Furthermore we can show that
\be
T_4(X_1,X_2) = 2\int_0^{\infty} ds \, \Ai(X_1+s)\Ai\hspace{0mm}'(X_2+s) + \Ai(X_2)\Ai(X_1)
=\frac{\partial}{\partial X_2}T_2(X_1,X_2)\ ,
\ee
and
\be
T_1(X_1,X_2) =2 \int_{X_2}^{X_1}dt \int_0^{\infty} ds \, \Ai(t+s)\Ai(X_2+s)-
\int_{X_2}^{X_1}dt\Ai(t) \int_{X_2}^{\infty} ds \, \Ai(s)\ ,
\ee
the latter of which can be verified by drawing a sketch of the $st$-plane showing the regions where there are contributions to the integrals, and comparing with the original form of $T_1$ in eq.\ (\ref{Tdefs}).
Our final result for the limiting matrix kernel thus agrees with the literature, see e.g.\ page 162 of \cite{Anderson}.

\sect{Some properties of an elementary integral}
\label{I_props}

Let us define the function for $x \in \RR$, $\tau > 0$ and integer $j \ge 0$
\be\label{Ij_redef}
I_j(x;\tau) \equiv \int_{-\infty}^{\infty} dt \, \sgn(x-t) w^{(1)}(t) H_j \left( \frac{t}{\sqrt{2\tau}} \right)
\ee
where, for real $x$,
\be
w^{(1)}(x) \equiv w^{(1)}(x;\tau) = \exp\left[ - \, \frac{x^2}{2(1+\tau)} \right],
\ee
and $H_j(x)$ is the physicists' Hermite polynomial of degree $j$.
To determine a useful recurrence relation involving the $I_j(x;\tau)$,
we begin by integrating eq.\ (\ref{Ij_redef}) by parts using eq.\ 8.952.1 of \cite{Grad}:
\begin{align}
I_j(x;\tau) &= \left[ \sgn(x-t) w^{(1)}(t) \frac{\sqrt{2\tau}}{2(j+1)} H_{j+1} \left( \frac{t}{\sqrt{2\tau}} \right) \right]_{-\infty}^{\infty} \nn \\
 & \qquad {} - \frac{\sqrt{2\tau}}{2(j+1)} \int_{-\infty}^{\infty} dt \, H_{j+1} \left( \frac{t}{\sqrt{2\tau}} \right) w^{(1)}(t) \left\{ -2\delta(x-t) + \sgn(x-t) \left( \frac{-t}{1+\tau} \right) \right\} \nn \\
 &= \frac{\sqrt{2\tau}}{j+1} H_{j+1} \left( \frac{x}{\sqrt{2\tau}} \right) w^{(1)}(x) + \frac{\sqrt{\tau}}{\sqrt{2}(1+\tau)(j+1)} \int_{-\infty}^{\infty} dt \, t \, \sgn(x-t) w^{(1)}(t) H_{j+1} \left( \frac{t}{\sqrt{2\tau}} \right).
\end{align}
From eq.\ 8.952.2 of \cite{Grad} we have the recurrence relation for Hermite polynomials
\be
t \, H_n\left( \frac{t}{\sqrt{2\tau}} \right) = \sqrt{2\tau} \left\{ \tfrac{1}{2}H_{n+1}\left( \frac{t}{\sqrt{2\tau}} \right) + n \, H_{n-1}\left( \frac{t}{\sqrt{2\tau}} \right) \right\},
\ee
and hence
\begin{align}
I_j(x;\tau) &= \frac{\sqrt{2\tau}}{j+1} H_{j+1} \left( \frac{x}{\sqrt{2\tau}} \right) w^{(1)}(x) + \frac{\tau}{(1+\tau)(j+1)} \left\{ \tfrac12 I_{j+2}(x;\tau) + (j+1)I_j(x;\tau) \right\}
\end{align}
which can be rearranged to give the following recurrence relation for the $I_j(x;\tau)$ themselves:
\begin{align}\label{Ij_recurrence_final}
I_j(x;\tau) &= \frac{\sqrt{2\tau}(1+\tau)}{j+1} H_{j+1} \left( \frac{x}{\sqrt{2\tau}} \right) w^{(1)}(x) + \frac{\tau}{2(j+1)} \, I_{j+2}(x;\tau).
\end{align}
We can evaluate the integral $I_j(x;\tau)$ by repeated application of eq.\ (\ref{Ij_recurrence_final}), to give for odd $j$:
\be\label{Ij_as_sum_of_Hermites_odd}
I_j(x;\tau) = - 2(1+\tau) \left( \frac{2}{\tau} \right)^{j/2} (j-1)!! \, w^{(1)}(x) \sum_{k=0}^{(j-1)/2} \frac{1}{(2k)!!} \left( \frac{\tau}{2} \right)^k H_{2k}\left(\frac{x}{\sqrt{2\tau}}\right),
\ee
and for even $j$:
\be\label{Ij_as_sum_of_Hermites}
I_j(x;\tau) = \left( \frac{2}{\tau} \right)^{j/2} (j-1)!! \left\{ I_0(x;\tau) - 2(1+\tau)w^{(1)}(x) \sum_{k=0}^{j/2-1} \frac{1}{(2k+1)!!} \left( \frac{\tau}{2} \right)^{k+1/2} H_{2k+1}\left(\frac{x}{\sqrt{2\tau}}\right) \right\}
\ee
where $I_0(x;\tau)$ can be written in terms of the error function as
\be\label{I0_explicit}
I_0(x;\tau) = \sqrt{2\pi(1+\tau)} \, \erf \left( \frac{x}{\sqrt{2(1+\tau)}} \right).
\ee

\sect{Hermitian limit of the $\beta=1$ matrix-kernel}
\label{Hermlimb1}

As with the $\beta=4$ case in Appendix \ref{Hermlimb4}, we verify that the Hermitian limit $\sigma\rightarrow 0$ of the matrix-kernel (and hence of all $k$-point correlation functions) agrees with known results in the literature.
Since we do not have the discontinuity associated with the $\beta=4$ ensemble,
here we can, in all cases, let $\sigma\rightarrow 0$ at fixed arguments (eigenvalues) $Z_j=X_j+iY_j$.
Indeed, note that no new Dirac delta functions appear in any of the following limits,
since these are already present in the non-Hermitian case, see eqs.\ (\ref{G4_lim_split}) and (\ref{W4_lim_split}).

It is useful to introduce the following general result for the pointwise limit of the complementary error function:
\be\label{erfc_herm_lim}
\lim_{\sigma\rightarrow 0} \erfc\left( \frac{|Y|}{\sigma} \right) = \delta_{Y0},
\ee
where, for $x,y \in \RR$, we define the Kr\"onecker delta function as
\be\label{kron_delta_def}
\delta_{xy} = \begin{cases} \,\, 1 & \text{if }x = y, \\ \,\, 0 & \text{otherwise}. \end{cases}
\ee

\noindent \textbf{(i)} We begin with $\hat{\mcK}_{\text{Ai}}^{(1)}(Z_1,Z_2)$, given in eq.\ (\ref{K1_lim_final}).   On expanding the $e^{\sigma^2 t}$ inside the integral, and then letting $\sigma\rightarrow 0$, this gives
\begin{align}
\lim_{\sigma\rightarrow 0} \hat{\mcK}_{\text{Ai}}^{(1)}(Z_1,Z_2) &= \frac{(X_1-X_2)}{4} \, \int_0^{\infty} dt \, t \, \Ai(X_1+t) \Ai(X_2+t) \, \delta_{Y_1 0} \delta_{Y_2 0}  \nn \\
 & = \frac{1}{4} \left\{ \int_0^{\infty} dt \, t \, \big[ \Ai\hspace{0mm}''(X_1+t) - t\,\Ai(X_1+t) \big] \, \Ai(X_2+t) - (X_1 \leftrightarrow X_2) \right\} \, \delta_{Y_1 0} \delta_{Y_2 0}\nn \\
 & = \frac{1}{4} \int_0^{\infty} dt \, t \, \big[ \Ai\hspace{0mm}''(X_1+t)\Ai(X_2+t) - (X_1 \leftrightarrow X_2)   \big] \, \delta_{Y_1 0} \delta_{Y_2 0}\nn \\
 & = \frac{1}{4} \left\{ - \int_0^{\infty} dt \, \Ai\hspace{0mm}'(X_1+t) \big[ t\,\Ai\hspace{0mm}'(X_2+t) + \Ai(X_2+t) \big] - (X_1 \leftrightarrow X_2) \right\} \, \delta_{Y_1 0} \delta_{Y_2 0}\nn \\
 & = \frac{1}{4} \left\{ - \int_0^{\infty} dt \, \Ai\hspace{0mm}'(X_1+t) \Ai(X_2+t) - (X_1 \leftrightarrow X_2) \right\} \, \delta_{Y_1 0} \delta_{Y_2 0}\nn \\
 &= \frac{1}{2} \left\{ \frac{\partial}{\partial X_2} \mcK_{\text{Ai,Herm}}^{(2)}(X_1,X_2) + \tfrac{1}{2} \, \Ai(X_1)\Ai(X_2) \right\} \, \delta_{Y_1 0} \delta_{Y_2 0},
\end{align}
where we used the fact that $\Ai''(X_1+t)=(X_1+t)\Ai(X_1+t)$ in the first step, cancelled terms, integrated by parts, cancelled more terms, and then integrated only the $X_1$-term by parts.   We also used the result eq.\ (\ref{erfc_herm_lim}) that the limit of the complementary error function involves the delta function defined in eq.\ (\ref{kron_delta_def}).
In the final line, $\mcK_{\text{Ai,Herm}}^{(2)}(X_1,X_2)=\int_0^{\infty} dt \, \Ai(X_1+t) \Ai(X_2+t)$ is the Hermitian limit of the $\beta=2$ kernel, see eq.\ (\ref{K2Herm_def}).

\noindent \textbf{(ii)} For $G_{\text{Ai}}^{\RR\,(1)}(Z_1,X_2)$, we take $\sigma\rightarrow 0$ in eq.\ (\ref{G_as_U1_plus_U23}). This gives
\be
- \lim_{\sigma\rightarrow\infty} G_{\text{Ai}}^{\RR\,(1)}(Z_1,X_2) = \left\{ \mcK_{\text{Ai,Herm}}^{(2)}(X_1,X_2) + \tfrac12 \, \Ai(X_1)\left( 1 - \int_0^{\infty} dt \, \Ai(X_2 + t) \right) \right\} \, \delta_{Y_1 0}.
\ee

\noindent \textbf{(iii)} For $G_{\text{Ai}}^{\CC\,(1)}(Z_1,Z_2)$, we have
\be
G_{\text{Ai}}^{\CC\,(1)}(Z_1,Z_2) = 2i \, \sgn(Y_2) \hat{\mcK}_{\text{Ai}}^{(1)}(Z_1,Z_2^*).
\ee
The pre-kernel $\hat{\mcK}_{\text{Ai}}^{(1)}(Z_1,Z_2^*)$ only has a non-zero limit when the arguments are both real
(i.e.\ when $Y_1=Y_2=0$).
But in this case, $\sgn(Y_2)=0$, and so $G_{\text{Ai}}^{\CC\,(1)}(Z_1,Z_2) \rightarrow 0$ for all arguments.

\noindent \textbf{(iv)} For $W_{\text{Ai}}^{\RR\RR\,(1)}(X_1,X_2)$, we simply set $\sigma = 0$ in eq.\ (\ref{WRR_limit_final}):
\begin{align}
- \lim_{\sigma\rightarrow 0} W_{\text{Ai}}^{\RR\RR\,(1)}(X_1,X_2) &=  \lim_{\sigma\rightarrow 0} \Big\{ - 2 A(X_1,X_2) + B(X_1)B(X_2) + B(X_2) - B(X_1) \Big\} \nn \\
&= 2 \left\{ \int_{X_2}^{X_1}\! dt \, \mcK_{\text{Ai,Herm}}^{(2)}(t,X_2) + \tfrac{1}{2} \left( \int_{X_2}^{X_1}\! dt \, \Ai(t)  \right)\! \left( 1 - \int_{X_2}^{\infty}\! dt \, \Ai(t) \right)\! \right\}.
\end{align}
The second step here can be proved with a few lines of easy manipulation.

\noindent \textbf{(v)} For $W_{\text{Ai}}^{\RR\CC\,(1)}(X_1,Z_2)$, we have
\be
W_{\text{Ai}}^{\RR\CC\,(1)}(X_1,Z_2) = 2i \, \sgn(Y_2) G_{\text{Ai}}^{\RR\,(1)}(Z_2^*,X_1).
\ee
The function $G_{\text{Ai}}^{\RR\,(1)}(Z_2^*,X_1)$ only has a non-zero limit when the first argument is real
(i.e.\ when $Y_2=0$).  But in this case, $\sgn(Y_2)=0$, and so $W_{\text{Ai}}^{\RR\CC\,(1)}(X_1,Z_2) \rightarrow 0$ for all arguments.

\noindent \textbf{(vi)} For $W_{\text{Ai}}^{\CC\CC\,(1)}(Z_1,Z_2)$, we have
\be
W_{\text{Ai}}^{\CC\CC\,(1)}(Z_1,Z_2) = -2i \, \sgn(Y_1) G_{\text{Ai}}^{\CC\,(1)}(Z_1^*,Z_2).
\ee
But $G_{\text{Ai}}^{\CC\,(1)}(Z_1^*,Z_2)\rightarrow 0$, and so $W_{\text{Ai}}^{\CC\CC\,(1)}(Z_1,Z_2)\rightarrow 0$ for all arguments.

\noindent \textbf{(vii)} The Hermitian limit of the bivariate weight function $\mcF_{\infty}^{(1)}(Z_1,Z_2)$ is trivial, since $\mcF_{\infty}^{(1)}(Z_1,Z_2)$ does not depend on $\sigma$.  We have simply
\be
\lim_{\sigma\rightarrow 0} \mcF_{\infty}^{(1)}(Z_1,Z_2) = 2i\delta^{(2)}(Z_1-Z_2^*)\sgn(Y_1) + \delta(Y_1)\delta(Y_2)\sgn(X_2-X_1).
\ee
In fact, we can drop the first term completely in the Hermitian limit, since it always gets multiplied by $\hat{\mcK}_{\text{Ai}}^{(1)}(Z_1,Z_2;\sigma=0) \propto \delta_{Y_1 0}$ when evaluating the Pfaffian to determine the correlation functions.

We expect the $\sigma\rightarrow 0$ limits to be consistent with those in the literature, e.g.\ on page 162 of \cite{Anderson}; our limit of $G_{\text{Ai}}^{(1)}(Z_1,Z_2)$ matches \cite{Anderson} precisely.
Our limit of $\hat{\mcK}_{\text{Ai}}^{(1)}(Z_1,Z_2)$ is one half of that given in \cite{Anderson},
but our limit of $W_{\text{Ai}}^{(1)}(Z_1,Z_2)$ is twice what is given in \cite{Anderson}.
It therefore follows that, once we evaluate the Pfaffian, we will get precisely the same $k$-point correlation functions for all values of $k$.


\begin{raggedright}	

\end{raggedright}

\end{document}